\documentclass[a4paper,12pt]{article}

\usepackage{amsfonts}
\usepackage{amsmath}
\usepackage{slashed}
\usepackage{mathrsfs}

\newcommand{\newsection}{    
\setcounter{equation}{0}\section}
\def\appendix#1{\addtocounter{section}{1}\setcounter{equation}{0}
\renewcommand{\thesection}{\Alph{section}}
\section*{Appendix \thesection\protect\indent \parbox[t]{11.15cm}{#1}}
\addcontentsline{toc}{section}{Appendix \thesection\ \ \ #1}}

\begin{document}
\begin{titlepage}
\begin{center}
\vspace*{-1.0cm}
\hfill DMUS--MP--15/06 \\

\vspace{2.0cm} {\Large \bf Geometry and supersymmetry of
heterotic warped flux  AdS backgrounds } \\[.2cm]

\vskip 2cm
S.~W.~ Beck$^1$,~J. B.  Gutowski$^2$ and G. Papadopoulos$^1$
\\
\vskip .6cm

\begin{small}
$^1$\textit{  Department of Mathematics, King's College London
\\
Strand, London WC2R 2LS, UK.\\
E-mail: samuel.beck@kcl.ac.uk
\\
E-mail: george.papadopoulos@kcl.ac.uk}
\end{small}\\*[.6cm]

\begin{small}
$^2$\textit{Department of Mathematics,
University of Surrey \\
Guildford, GU2 7XH, UK. \\
Email: j.gutowski@surrey.ac.uk}
\end{small}\\*[.6cm]

\end{center}

\vskip 3.5 cm
\begin{abstract}

 We classify the geometries of the most general warped, flux  AdS backgrounds of heterotic supergravity up to two loop order in sigma model perturbation theory.
 We show  under some mild assumptions that there are no $AdS_n$ backgrounds with $n\not=3$. Moreover the warp factor  of AdS$_3$ backgrounds is constant,    the geometry  is a product
 $AdS_3\times M^7$ and such solutions preserve, 2, 4, 6 and 8 supersymmetries. The geometry of $M^7$ has been specified in all cases.  For 2 supersymmetries, it has been found
 that $M^7$ admits a suitably restricted $G_2$ structure. For 4 supersymmetries,  $M^7$ has an $SU(3)$ structure and can be described locally as a circle fibration over a 6-dimensional KT manifold. For 6 and 8 supersymmetries, $M^7$ has an $SU(2)$ structure and can be described locally as a $S^3$ fibration over a 4-dimensional manifold which  either has an  anti-self dual Weyl tensor   or
  a hyper-K\"ahler structure, respectively. We also demonstrate a new  Lichnerowicz type theorem in the presence of  $\alpha'$ corrections.

\end{abstract}

\end{titlepage}


\section{Introduction}

For over 30 years, AdS backgrounds have found widespread applications  originally in supergravity \cite{FR}  and more recently in
string theory compactifications, and  in AdS/CFT, see reviews \cite{duff, grana, maldacena}. Despite many developments, the geometry of all such backgrounds has not been specified and
their classification remains an open problem, for some selected publications see \cite{romans}-\cite{nunez}.
Only recently, the fractions of supersymmetry preserved by  flux, warped AdS backgrounds of 11-dimensional
and type II supergravities have been specified in \cite{mads, iibads, iiaads} without making any assumptions on the form of the fields and that of the Killing spinors apart from imposing the symmetries of AdS space on the former. The computation of these fractions has also uncovered a new class of Lichnerowicz  type theorems.

To make progress towards the classification of AdS backgrounds, it is instructive to have a paradigm where all the issues that arise can be
worked out in a simpler setting than that of the 11-dimensional
and type II supergravities. Such theories are the common sector of type II supergravities and the heterotic supergravity. One simplification that occurs in
the common sector and heterotic supergravities is
that all solutions of the KSEs  are known and the geometry of the backgrounds has been determined \cite{hetclas}. In particular,  AdS$_3$ configurations arise naturally
 in the context of classification of  supersymmetric heterotic backgrounds \cite{hetclas} and their geometry has also been examined\footnote{Our description of geometry differs
from that of \cite{ohta} specially for backgrounds preserving more than 2 supersymmetries.} in \cite{ohta}.  The AdS backgrounds are also special cases of the heterotic horizons
in \cite{hethor} though the latter has been explored only for a closed 3-form field strength $H$. In addition, there is  a no-go theorem for AdS$_4$ heterotic backgrounds that has been
established  in \cite{ads4nogo} without assuming supersymmetry but imposing smoothness on the fields and compactness of the transverse 6-dimensional space.

The $\alpha'$ corrections to heterotic  backgrounds can be investigated in the perturbative or sigma model  approach, see eg \cite{sbook} and references within. In this, the field equations of the theory are the vanishing
conditions of the sigma model beta functions which are expressed as a series in $\alpha'$, the string tension. Furthermore, the Bianchi identity  of the 3-form field strength gets corrected as part
 of a mechanism which cancels the chiral reparameterization and gauge anomalies of the 2-dimensional sigma model.  This is referred to as the anomaly cancelation mechanism. As a result $H$ is not closed any longer at one and higher loops. Instead, $H$ at one-loop obeys the ``anomalous Bianchi identity'' in which $dH$
is equal to the difference between the Pontryagin form of the gravitational sector and that of the gauge sector of the heterotic theory.   After the modification of the
 Bianchi identity of $H$ at one loop,  consistency of the theory
requires the inclusion of the two loop correction to the field equations. Generic heterotic backgrounds receive $\alpha'$ corrections
 to all loop orders.

There is also interest in the differential systems which arise from truncating the theory up to two loops in perturbation theory. There are at least two approaches to these. One
is to consider the systems which arise from the KSEs and the anomalous Bianchi identity, and the other is to in addition consider  the
field equations up to and including two loops. The difference is that in the former case there are more options in the choice of connections which enter in the Pontryagin forms of the
anomalous Bianchi identity. In the latter case,  consistency of the differential system requires that these
connections satisfy an instanton type of condition.
Solutions to such systems have been found sometime ago in \cite{opfermann} in connection with HKT geometry \cite{hkt} on homogenous spaces. More recently attention has been
 centered on  the solutions \cite{becker, fu, ivanovb} of the  Strominger system which is associated with ${\mathbb{R}}^{3,1}$ compactifications of the heterotic string \cite{strominger, hull} and  hermitian geometry.

 AdS spaces can be seen as backgrounds in the perturbative approach to the heterotic string, and also as solutions to the differential system that arises after the theory is truncated to two loops.
 Because of this, both points of view will be explored, though the former is suitable for applications to the heterotic string. There are some differences in the analysis between the two
 cases. One difference is that in  perturbation theory one begins with a background which satisfies $dH=0$ and then corrects it order by order
 in $\alpha'$. This is because one starts the perturbation theory from a world-sheet classical action which has a B-field coupling to the string. The existence of such zeroth order backgrounds puts a restriction on the theory.
 It is known for example that there are no-go theorems for such backgrounds that preserve
  some of the spacetime superymmetry \cite{ivanovgp}. If the zeroth order background exists and the perturbation theory can be set up,  then typically the question that arises
 is whether there is a renormalzation scheme such that  the symmetries of the zeroth order background get maintained to all orders in
 perturbation theory.  Such computations for world-sheet supersymmetries can be found in   \cite{howe40} and for spacetime supersymmetries in \cite{tsimpis, lu}, and more recently in \cite{beckerb}.

On the other hand if the theory is truncated to two loops, $\alpha'$ is not treated as an expansion parameter but rather as a constant. As the anomalous Bianchi identity
and the field equations do not have an explicit dependence on the B-field, the fields that are restricted by the   differential system are the metric and the fundamental forms of the G-structure as
typically $H$ is expressed in terms of them.
Therefore, the requirement of existence of $dH=0$ configurations does not arise in the investigation of the differential system.

In this paper, we shall classify the geometry of supersymmetric warped, flux AdS backgrounds up to two loops in both heterotic perturbation theory and in the differential system
that arises after truncating the higher than two loop contributions. We find  in both cases that  AdS$_3$ heterotic supergravity  backgrounds admit 2, 4, 6 and 8 supersymmetries.
We show that the geometry is always a product $AdS_3\times M^7$, ie the warp factor is constant. Furthermore, the two spacetime supersymmetries restrict $M^7$
to have a conformally balanced  $G_2$ structure compatible with a connection with skew-symmetric torsion. Four spacetime supersymmetries restrict $M^7$ to have a $SU(3)$
structure compatible with a   connection with skew-symmetric torsion. In addition $M^7$ is locally a circle fibration over a conformally balanced  KT manifold with a $U(3)$
structure and holomorphic canonical bundle. Six spacetime supersymmetries restrict $M^7$ to have a $SU(2)$ structure.  Moreover $M^7$ is locally a $S^3$ fibration
over a 4-dimensional manifold $B^4$ which has vanishing self-dual Weyl tensor. The fibres $S^3$ twist over $B^4$ with an $\mathfrak{su}(2)$ connection whose self-dual
part is specified by a quaternion K\"ahler structure on $B^4$ and its anti-self-dual part is unrestricted. Eight spacetime supersymmetries
restrict $M^7$ to have a $SU(2)$  structure. Now $M^7$ is locally a $S^3$ fibration over a 4-dimensional hyper-K\"ahler manifold $B^4$. The $S^3$ fiber
is twisted with respect to an anti-self-dual  $\mathfrak{su}(2)$ connection.

We also show that there are no AdS$_n$ backgrounds for $n\not=3$. For $n>3$, this result has been established up to two loops in both perturbation theory
and in the differential system approaches without making any further assumptions on the backgrounds apart from supersymmetry.  For $n=2$ this result has been established for backgrounds with $dH=0$ and after assuming the application of the maximum principle\footnote{AdS$_3$ can be written as a warped product of AdS$_2$
\cite{stromingerads2} and so there exist heterotic AdS$_2$ backgrounds.  However they  do not satisfy the maximum principle.}, and follows from the results of \cite{iibhorizon}.

In addition we prove a new Lichnerowicz type theorem in the presence of  $\alpha'$ corrections. This generalizes the theorem we have shown for
$dH=0$ configurations in \cite{iibads} and \cite{iiaads}. We find that after choosing an appropriate Dirac operator that couples to fluxes on $M^7$, all the zero
modes of this Dirac operator are Killing spinors of the gravitino,  dilatino and, in the perturbation theory, the gaugino KSEs.  Furthermore, the validity of the theorem requires an
appropriate choice for a connection of $M^7$ and that of the gauge sector which contribute to the higher order $\alpha'$ terms of field equations and anomalous Bianchi identity.

This paper is organized as follows. In section 2, we investigate the geometry of AdS$_3$ backgrounds provided $dH=0$ and prove a Lichnerowicz type theorem.  In section 3,  we investigate the geometry of AdS$_3$ backgrounds provided $dH\not=0$ and establish a Lichnerowicz type theorem for backgrounds with $\alpha'$ corrections. In section 4, we demonstrate that there are no
AdS$_n$ backgrounds for $n\not=3$. In appendices A and B, we summarize our notation and give some formulae needed for the proof
of the Lichnerowicz type theorem, respectively. In appendix C, we derive the dilatino field equation from the other field equations of the theory. In appendix D, we examine
the geometry of of AdS$_3$ backgrounds from another choice of Killing spinors and in appendix E, we give a component based proof on the non-existence
of AdS$_n$, $n\not=3$, backgrounds.

\newsection{AdS$_3$ backgrounds with $dH=0$}

The investigation of AdS$_3$ backgrounds will be separated into two cases depending on whether $dH$ vanishes or not. For the common sector
of type II supergravities as well as that for the heterotic string with the standard embedding which leads to the vanishing of the chiral anomaly, one has $dH=0$. Furthermore $dH=0$ at zeroth order in the $\alpha'$ expansion in the sigma model approach to the heterotic string.  However, in the latter case $dH\not=0$ to one and higher loops.  For applications to the common sector, it is understood that we consider only one of the two chiral copies of the KSEs.

\subsection{Fields, Field Equations and Bianchi Identities}
\label{fads3}

The most general  metric and NS-NS 3-form flux of warped AdS$_3$ backgrounds which are invariant under the action of the $\mathfrak{sl}(2,{\mathbb{R}})\oplus \mathfrak{sl}(2,{\mathbb{R}})$
symmetry algebra of AdS$_3$ are
\begin{eqnarray}
 ds^2 &=& 2 \mathbf{e}^+ \mathbf{e}^- + A^2 dz^2 + ds^2 \left( M^7 \right) ,
 \cr
  H &=& A X e^+ \wedge e^- \wedge dz + G ,
  \label{fansatz}
 \end{eqnarray}
 where we have introduced the frame
 \begin{eqnarray}
 \mathbf{e}^+ &=& du \qquad \mathbf{e}^- = dr - \frac{2r}{\ell} dz - 2r d \ln A~,
 \end{eqnarray}
 $u, v,$ and $z$ are the AdS$_3$ coordinates, $\ell$ is the AdS radius, and $A$ is the warp factor. For more details on this parametrization of AdS$_3$ backgrounds see \cite{adsfbh}.
 Furthermore, one finds that the dilaton, $\Phi$, and the warp factor, $A$, and $G$ depend only on the $M^7$ coordinates.  In addition $X$ and $A$, and $G$ are functions, and a 3-form on $M^7$, respectively.

 The heterotic theory has in addition  a 2-form gauge field $F$ with gauge group a subgroup of $E_8\times E_8$ or $SO(32)/{\mathbb{Z}}_2$ that is associated with the gauge sector.  One way to impose the
 symmetries of AdS$_3$ on $F$ is to take  $F$ to be the curvature of a connection on $M^7$ that depends only on the coordinates of $M^7$. Alternatively, the gaugino\footnote{From now on we assume that the gaugino KSE has the same Killing spinors as the gravitino KSE, see \cite{hetclas} for a justification.} KSE for the backgrounds that we shall be considering implies
 that $F$ vanishes along the AdS$_3$ directions and that the Lie derivative of $F$ along the isometries of AdS$_3$ vanishes as well  up to  gauge transformations. These in particular imply
 that the associated Pontryagin forms vanish along AdS$_3$ and depend only on the coordinates of $M^7$. Either results are sufficient for the analysis that will follow.

 So far, we have not imposed the Bianchi identity on $H$ and (\ref{fansatz}) applies equally to backgrounds regardless on whether $dH$ vanishes or not.
 However imposing now the Bianchi identity, $dH=0$, one finds
\begin{eqnarray}
 d(A^3 X) = 0~,~~~
 dG = 0~.
 \label{bid}
\end{eqnarray}
The field equations for the dilatino and 2-form gauge potential can be expressed as
\begin{eqnarray}
 \nabla^2 \Phi &=& -3 A^{-1} \partial_i A \partial^i \Phi + 2 (d\Phi)^2 - \frac{1}{12} G^2 + \frac{1}{2} X^2~,
 \cr
 \nabla^k G_{i j k} &=& -3 A^{-1} \partial^k A G_{i j k} + 2 \partial^k \Phi G_{i j k}~,
\end{eqnarray}
where $i,j,k=1,\dots, 7$.
Moreover, the AdS component of the Einstein equation reads
\begin{equation}
 \nabla^2 \ln A = -\frac{2}{\ell^2} A^{-2} - 3 A^{-2} (dA)^2 + 2 A^{-1} \partial_i A \partial^i \Phi + \frac{1}{2} X^2 ,
\end{equation}
and the $M^7$ components are
\begin{equation}
 R^{(7)}_{i j} = 3 \nabla_i \nabla_j \ln A + 3 A^{-2} \partial_i A \partial_j A + \frac{1}{4} G_{i k_1 k_2} G_j{}^{k_1 k_2} - 2 \nabla_i \nabla_j \Phi ,
\end{equation}
where $\nabla$ is the Levi-Civita connection on $M^7$ and $R^{(7)}_{i j}$ is its Ricci tensor. The Ricci scalar curvature of $M^7$ can be expressed
\begin{eqnarray}
 R^{(7)} &=& 3 \nabla^2 \ln A + 3 A^{-2} (dA)^2 + \frac{1}{4} G^2 - 2 \nabla^2 \Phi
 \cr
 &=& -\frac{6}{\ell^2} A^{-2} - 6 A^{-2} (dA)^2 + \frac{5}{12} G^2 + \frac{1}{2} X^2 + 12 A^{-1} \partial_i A \partial^i \Phi - 4 (d\Phi)^2 .
\end{eqnarray}
This formula for $A$ constant will be used later in the proof of a Lichnerowicz  type theorem.

\subsection{Solution of KSEs along AdS$_3$}

The heterotic gravitino and dilatino KSEs are
\begin{eqnarray}
 \nabla_M \epsilon - \frac{1}{8} \slashed{H}_M \epsilon = 0+{\cal O}(\alpha'^2)~,~~~
 \big(\slashed{\partial}\Phi-{1\over12} \slashed{H}\big)\epsilon=0+{\cal O}(\alpha'^2)~.
 \label{ksealpha1}
\end{eqnarray}
Therefore, the form of the two KSEs remains the same up to two and possibly higher loops. The gaugino KSE does not contribute
in the investigation of backgrounds with $dH=0$  and so it is not included.

First let us focus on the gravitino KSE.  The gravitino KSE along the  AdS$_3$ directions reads
\begin{eqnarray}
 \partial_u \epsilon_\pm + A^{-1} \Gamma_{+ z} \left( \ell^{-1} - \Xi_- \right) \epsilon_\mp&=&0~,
 \cr
 \partial_r \epsilon_\pm - A^{-1} \Gamma_{- z} \Xi_+ \epsilon_\mp&=&0~,
 \cr
 \partial_z \epsilon_\pm - \Xi_\pm \epsilon_\pm + \frac{2r}{\ell} A^{-1} \Gamma_{- z} \Xi_+ \epsilon_\mp&=&0~,
 \label{ksex}
\end{eqnarray}
where
\begin{equation}
 \Xi_\pm = \mp \frac{1}{2 \ell} + \frac{1}{2} \slashed{\partial} A \Gamma_z \mp \frac{1}{4} A X ,
\end{equation}
and $\Gamma_\pm\epsilon_\pm=0$.
Furthermore, using the relations
\begin{eqnarray}
 \Xi_\pm \Gamma_{z +} + \Gamma_{z +} \Xi_\mp = 0~,~~~
 \Xi_\pm \Gamma_{z -} + \Gamma_{z -} \Xi_\mp = 0~,
\end{eqnarray}
we find that there is only one independent integrability condition
\begin{eqnarray}
  \left( {\Xi_\pm}^2 \pm \frac{1}{\ell} \Xi_\pm \right) \epsilon_\pm= \left( -\frac{1}{4 \ell^2} - \frac{1}{4} (dA)^2 \mp \frac{1}{4} A X \slashed{\partial} A \Gamma_z + \frac{1}{16} A^2 X^2 \right) \epsilon_\pm =0~.
\end{eqnarray}
As the Clifford algebra operator $\slashed{\partial} A \Gamma_z$ does not have real eigenvalues,  the above integrability condition for  $\ell < \infty$ can  be satisfied provided that
\begin{equation} \label{eq:simplified_AdS_integrability}
 dA=0~,~~~-\frac{1}{4 \ell^2} + \frac{1}{16} A^2 X^2 = 0~.
\end{equation}
Thus the warp factor $A$ is constant. The second equation above  also  implies that the component $X$ of $H$ along AdS$_3$ is  constant.
Furthermore, one can write
\begin{equation}
 \Xi_\pm = \mp \frac{1 + c_1}{2 \ell}~,
 \label{xie}
\end{equation}
where $c_1=\frac{\ell}{2}AX=\pm 1$ as implied by \eqref{eq:simplified_AdS_integrability}.

The KSEs (\ref{ksex}) can be integrated to find
\begin{eqnarray}
\epsilon=\epsilon_++\epsilon_-=\sigma_++ e^{-{z\over\ell}} \tau_++\sigma_-+ e^{{z\over\ell}} \tau_--\ell^{-1} u A^{-1} \Gamma_{+z} \sigma_--\ell^{-1} r A^{-1} e^{-{z\over\ell}}\Gamma_{-z} \tau_+~,
\end{eqnarray}
provided that
\begin{equation}
 \Xi_\pm \sigma_\pm = 0 \qquad \Xi_\pm \tau_\pm = \mp \frac{1}{\ell} \tau_\pm~.
 \label{xicon}
\end{equation}
It is understood that the dependence of $\epsilon$ on the AdS$_3$ coordinates is given explicitly while $\tau_\pm$ and $\sigma_\pm$ depend only on the coordinates of $M^7$.

It is clear from (\ref{xicon}) that there are two solutions to the above conditions. If $c_1=1$, (\ref{xicon}) implies that $\sigma_\pm=0$. In turn
the Killing spinor is
\begin{eqnarray}
\epsilon=\epsilon_++\epsilon_-= e^{-{z\over\ell}} \tau_++ e^{{z\over\ell}} \tau_--\ell^{-1} r A^{-1} e^{-{z\over\ell}}\Gamma_{-z} \tau_+~.
\end{eqnarray}
Alternatively if $c_1=-1$, (\ref{xicon}) gives $\tau_\pm=0$ and the Killing spinor is
\begin{eqnarray}
\epsilon=\epsilon_++\epsilon_-=\sigma_++\sigma_--\ell^{-1} u A^{-1} \Gamma_{+z} \sigma_-~.
\end{eqnarray}
Therefore depending on the sign of $AX$, which coincides in the sign of the contribution volume form of AdS$_3$  in $H$, there are two distinct cases to consider.

In order to interpret the two cases that arise, note that AdS$_3$ can be identified, up to a discrete identification, with the group manifold $SL(2,{\mathbb{R}})$.  As such it is parallelizable with respect
to either left or right actions of $SL(2,{\mathbb{R}})$.  The two associated connections differ by the sign of their torsion term which in turn  is given by the structure constants of the
$\mathfrak{sl}(2,{\mathbb{R}})$.  Of course the associated 3-form coincides with the bi-invariant volume form of AdS$_3$.

To treat both cases symmetrically, we introduce $ \mathbb{B}^{(\pm)} $ which is  equal to $ \Xi_\pm $ when it acts on $ \sigma_\pm $ and equal to $ \Xi_\pm \pm \frac{1}{\ell} $ when it acts on $ \tau_\pm $. The integrability conditions are  then succinctly expressed as $ \mathbb{B}^{( \pm )} \chi_\pm = 0 $, $ \chi_\pm = \sigma_\pm, \tau_\pm $, where
\begin{equation}
 \mathbb{B}^{(\pm)} = \mp \frac{c_1 + c_2}{2 \ell}~,
\end{equation}
with $c_2=1$ when $\chi_\pm=\sigma_\pm$ and with $c_2=-1$ when $\chi_\pm=\tau_\pm$.

The remaining KSEs on $M^7$ can now be expressed as
\begin{equation}
 \nabla_i^{(\pm)} \chi_\pm = 0, \qquad \mathcal{A}^{(\pm)} \chi_\pm = 0, \qquad  \mathbb{B}^{(\pm)} \chi_\pm = 0~,
 \label{rkse3}
\end{equation}
where
\begin{eqnarray}
\nabla^{( \pm )}_i  = \nabla_i + \Psi^{( \pm )}_i~,~~~\Psi^{( \pm )}_i = -\frac{1}{8} \slashed{G}_i~,
\end{eqnarray}
is a metric connection with skew-symmetric torsion $G$ associated with the gravitino KSE, and
\begin{equation}
 \mathcal{A}^{(\pm)} = \slashed{\partial} \Phi \pm \frac{c_1}{\ell} A^{-1} \Gamma_z - \frac{1}{12} \slashed{G}~,
\end{equation}
is associated with the dilatino KSE, and $\mathbb{B}^{(\pm)}$ should be thought as a projector which restricts the first
two equations on either $\sigma_\pm$ or $\tau_\pm$ spinors.

For the investigation of the geometry of these backgrounds it suffices to consider only the $\tau_+$ or the $\sigma_+$ spinors.
This is because, if $\chi_-$ is a solution to the above KSEs, then $\chi_+=A^{-1} \Gamma_{+z} \chi_-$ is also a solution, and vice versa,
if $\chi_+$ is a solution, then $\chi_-=A\Gamma_{-z} \chi_+$ is also a solution. Incidentally, this also implies that the number
of supersymmetries preserved by AdS$_3$ backgrounds is always even.  Furthermore, it suffices to investigate the geometry of these backgrounds as  described
by the  $\sigma_+$ spinors. As we have mentioned, the $\tau_+$ spinors arise on choosing the other parallelization for AdS$_3$ and it can be treated
symmetrically, see also appendix D.

\subsection{Geometry}
\label{g2geom}

If the solution of the KSEs is determined by the $\sigma_\pm$ spinors, the investigation of the geometry of $M^7$ can be done as a special case of that of heterotic horizons in \cite{hethor} which utilized the classification results
of \cite{hetclas}.  To see this first note that $h=-{2\over\ell} dz$ and so  the constant $k$ which enters in the description
of geometry for the heterotic horizons is
\begin{equation}
 k^2=h^2= 4 A^{-2} \ell^{-2}~.
 \end{equation}
 Next observe that as $\sigma_+$ and $\sigma_-$ are linearly independent, there are two Killing spinors given by
 \begin{eqnarray}
 \epsilon^1=\sigma_+~,~~~\epsilon^2=\sigma_--\ell^{-1} u A^{-1} \Gamma_{+z} \sigma_-~.
 \end{eqnarray}
 Setting now $\sigma_-=A \Gamma_{-z} \sigma_+$ and after rescalling the second spinor with the non-vanishing constant $-2\ell^{-1} A^{-2}$, we find that
 the two spinors can be rewritten as
\begin{eqnarray}
 \epsilon^1=\sigma_+~,~~~\epsilon^2=-k^2 u  \sigma_++\Gamma_-\slashed{h} \sigma_+~.
 \label{het2ks}
\end{eqnarray}
These are precisely the spinors that appear in the context of heterotic horizons,   see \cite{hethor} for a detailed description of the geometry of $M^7$ including
the emergence of the (left)  $\mathfrak{sl}(2,{\mathbb{R}})$ symmetry of AdS$_3$ backgrounds as generated by the 1-form Killing spinor bi-linears.  Briefly, $M^7$ admits a $G_2$ structure
compatible with a metric connection $\hat\nabla$ with skew-symmetric torsion, $\hat\nabla_i X^j=\nabla_i X^j+{1\over2} G^j{}_{ik} X^k$. Furthermore all field equations
and KSEs are implied provided \cite{hethor} that
\begin{eqnarray}
d\big(e^{2\Phi} \star_7 \varphi\big)=0~,~~~dG=0~,
\label{g2con}
\end{eqnarray}
where
\begin{eqnarray}
G=k\varphi+ e^{2\Phi} \star_7 d(e^{-2\Phi} \varphi)~,
\label{gg2flux}
\end{eqnarray}
and $\varphi$ is the fundamental form of the $G_2$ structure.  The first condition in (\ref{g2con}) is required for the existence of a $G_2$ structure on $M^7$ compatible
with a metric connection with skew-symmetric torsion \cite{ivanovy} and the second condition is the Bianchi identity (\ref{bid}). The dilatino KSE implies two conditions,
one of which is that the $G_2$ structure on $M^7$ must be conformally balanced, $\theta_\varphi=2d\Phi$, both of which have been incorporated in the expression for $G$,
where $\theta_\varphi$ is the Lee form of $\varphi$.
The conditions (\ref{g2con}) are simpler
than those that have appeared for heterotic horizons, because for AdS$_3$ backgrounds $dh=0$.

\subsection{Geometry of AdS$_3$ backgrounds with extended supersymmetry}
\label{extendsusy}

We have shown that AdS$_3$ backgrounds always preserve an even number of supersymmetries. Furthermore, from the counting of supersymmetries for  heterotic horizons \cite{hethor}, one concludes that  AdS$_3$ backgrounds preserve 2, 4, 6 and 8 supersymmetries. In addition, AdS$_3$ backgrounds that
preserve 8 supersymmetries and for which $M^7$ is compact are locally isometric to either $AdS_3\times S^3\times T^4$ or to $AdS_3\times S^3\times K_3$.  Again we shall
not give the details of the proof for these results. However, we shall state the key formulae that arise in the investigation of the geometry for each case
as they have some differences from those of the heterotic horizons.

\subsubsection{Four supersymmetries}
Let us first consider the AdS$_3$ backgrounds with 4 supersymmetries.  The two additional spinors can be written as
\begin{eqnarray}
 \epsilon^3=\sigma^2_+~,~~~\epsilon^4=-k^2 u  \sigma^2_++\Gamma_-\slashed{h} \sigma^2_+~,
\end{eqnarray}
where $\sigma^2_+$ is linearly independent from $\sigma_+^1=\sigma_+$ in (\ref{het2ks}). In fact it can be shown that the normal form for these
spinors up to the action of $Spin(7)$ can be chosen as $\sigma^1_+=1+e_{1234}$ and $\sigma^2_+=i(1-e_{1234})$.  The isotropy group of all four  spinors is
$SU(3)$.  Therefore $M^7$ is a Riemannian manifold equipped with metric $ds^2_{(7)}$ and a 3-form $G$. Furthermore,  the metric connection $\hat\nabla$ with  skew-symmetric torsion $G$ is  compatible with  an $SU(3)$ structure. The KSEs
restrict this structure on $M^7$ further.   In particular, the $SU(3)$ structure on  $M^7$ is associated with
1-form $\xi$, 2-form $\omega$, and  (3,0)-form $\chi$  spinor bilinears such that
\begin{eqnarray}
i_\xi \omega=0~,~~~{\cal L}_\xi \omega=0~,~~~i_\xi \chi=0~,~~~{\cal L}_\xi\chi =i k \chi~,
\label{funcon}
\end{eqnarray}
where $\omega$ and $\chi$ are the fundamental forms of an $SU(3)$ structure  in the directions transverse to $\xi$. All these forms are  $\hat\nabla$-parallel, $\hat\nabla \xi=\hat\nabla \omega=\hat\nabla\chi=0$.  In particular $\hat\nabla \xi=0$ implies
that $\xi$ is Killing and that  $i_\xi G=k^{-1} dw$, where $w(\xi)=k$. As $G$ is closed ${\cal L}_\xi G=0$.  The dilaton $\Phi$ is also invariant under $\xi$. The full set of conditions on $\xi$, $\omega$ and $\chi$ can be found in \cite{hethor}.

The solution of these conditions\footnote{In fact with the data provided $M^7$ admits a normal almost contact structure which however is further restricted.} implies that  $M^7$ can be locally constructed
as a circle fibration on  a conformally balanced\footnote{$\theta_\omega$ is the Lee form of $B^6$.}, $\theta_\omega=2d\Phi$, KT manifold $B^6$ with Hermitian form $\omega$, where the tangent space of the circle fibre is spanned by $\xi$.  The  canonical bundle of $B^6$ admits a connection\footnote{There some differences in the notation of this paper with that of \cite{hethor}. For example $w$ is denoted in \cite{hethor} with $\ell$. We have made this change   because here we have denoted by $\ell$ the radius of
 AdS.} $\lambda=k^{-1} w$,  such that
\begin{eqnarray}
dw^{(2,0)}=0~,~~~dw_{ij} \omega^{ij}=-2k^2~,
\end{eqnarray}
$i, j=1,2,\dots, 6$, i.e. the canonical bundle is holomorphic and the connection satisfies the Hermitian-Einstein instanton condition, and in addition
\begin{eqnarray}
\hat\rho_{(6)}=dw~,~~~k^{-2} dw\wedge dw+d G_{(6)}=0~,
\end{eqnarray}
where
\begin{eqnarray}
\hat\rho^{(6)}_{ij}={1\over2} \hat R^{(6)}{}_{ij}{}^k{}_m I^m{}_k~,
\end{eqnarray}
 is the curvature of the canonical bundle induced from the connection with torsion $G_{(6)}=-i_I d\omega$ on $B^6$,  and $I$ is the complex structure of $B^6$.  The first condition is required for $M^7$ to admit an $SU(3)$ structure compatible with the connection with skew-symmetric torsion $G$ and
the second condition is required by the Bianchi identity (\ref{bid}). Note that $B^6$ has a $U(3)$ rather than an $SU(3)$ structure compatible with a connection with skew-symmetric
torsion.  This is because the (3,0)-form $\chi$ is not invariant under the action of $\xi$ (\ref{funcon}).

The metric and torsion on $M^7$ are given from those of $B^6$ as
\begin{eqnarray}
ds^2(M^7)=k^{-2} w^2+ ds^2(B^6)~,~~~G= k^{-2} w\wedge dw+ G_{(6)}~.
\label{6from7}
\end{eqnarray}
This summarizes the geometry for the AdS$_3$ backgrounds preserving four supersymmetries. Solutions can be constructed using the techniques developed in \cite{hethor}
to find solutions for heterotic horizons.

\subsubsection{Six supersymmetries}

Next let us turn to AdS$_3$ backgrounds preserving 6 supersymmetries. For these $M^7$ admits an $SU(2)$ structure compatible with $\hat\nabla$. Furthermore, $M^7$ can be constructed locally as a $SU(2)=S^3$ fibration over
a 4-dimensional manifold $B^4$ whose self-dual part of the Weyl tensor vanishes. $SU(2)$ twists over $B^4$ with respect to a (principal bundle) connection $\lambda$ which has
curvature ${\cal F}^{r'}$ such that the self-dual part satisfies
\begin{eqnarray}
({\cal F}^{\mathrm{sd}})^{r'}={k\over 4} \omega^{r'}~,
\end{eqnarray}
 where $ \omega^{r'}$ are the almost Hermitian forms of a quaternionic K\"ahler structure on $B^4$. The anti-self dual part of ${\cal F}$, ${\cal F}^{\mathrm{ad}}$, is not restricted by the KSEs.
The dilaton depends only on the coordinates of $B^4$.
The metric and $G$ on $M^7$ are given by
\begin{eqnarray}
ds^2(M^7)=\delta_{r's'} \lambda^{r'} \lambda^{s'} + e^{2\Phi} d\mathring s^2(B^4)~,~~~G=CS(\lambda)-\mathring{\star} d e^{2\Phi}~,
\label{mtor6}
\end{eqnarray}
where $CS$ is the Chern-Simons\footnote{Note that if ${\cal F}=0$, $CS(\lambda)$ is proportional
to the volume of $S^3$.}  form of $\lambda$.
The only condition that remains to be solved to find solutions is
\begin{eqnarray}
\mathring{\nabla}^2 e^{2\Phi}=-{1\over2} ({\cal F}^{\mathrm{ad}})^2+{3\over8} k^2 e^{4\Phi}~,
\label{fcon6}
\end{eqnarray}
where the inner products are taken with respect to the $d\mathring s^2$ metric.
For more details on the geometry of such backgrounds see \cite{hethor}.

\subsubsection{Eight supersymmetries}

Next let us turn to the AdS$_3$ backgrounds preserving 8 supersymmetries.  The description of the geometry is as that of the backgrounds above preserving 6 supersymmetries.
The only differences are that $B^4$ must be a hyper-K\"ahler manifold with respect to the $d\mathring s^2(B^4)$ metric, and that ${\cal F}^{\mathrm{sd}}=0$. The metric and 3-form $G$ of $M^7$ are given
as in (\ref{mtor6}) but now we have that
\begin{eqnarray}
\mathring{\nabla}^2 e^{2\Phi}=-{1\over2} ({\cal F}^{\mathrm{ad}})^2~,
\end{eqnarray}
instead of (\ref{fcon6}).  If $B^4$ is compact, a partial integration argument reveals tha ${\cal F}^{\mathrm{ad}}=0$ and so the only regular solutions, up to discrete identifications, are
$AdS_3\times S^3\times K_3$ and
$AdS_3\times S^3\times T^4$. If $B^4$ is not compact, there are many  smooth solutions, see \cite{hethalf}.

\subsection{Lichnerowicz type theorem on $ \sigma_+ $, $ \tau_+ $}

The Killing spinors of AdS$_3$ backgrounds (\ref{rkse3}) can be identified with the zero modes of a suitable Dirac-like operator coupled to fluxes on $M^7$, and vice versa.
This provides  a new example of a  Lichnerowicz type  theorem for connections whose holonomy is not in a Spin group.
This result is analogous to others that have been established for  AdS backgrounds in 11-dimensional and type II supergravities \cite{mads, iibads, iiaads}.
However, there are some differences. One is that the spinor representation in the heterotic case is different from that of the previous
mentioned theories. There are also some subtle issues
associated with the modification of the Lichnerowicz type of theorem
in the presence of $\alpha'$ corrections, which we shall consider
in further detail in the next section.

To begin, let us  first suppress the $\alpha'$ corrections, and take $dH=0$.
The Lichnerowicz type of theorem with $\alpha'$ corrections will be investigated later. We define the modified
 gravitino Killing spinor operator,
\begin{equation}
 \hat{\nabla}^{(+, q_1, q_2)}_{i} = \nabla^{(+)}_{i} + \Gamma_i  \mathbb{A}^{( +, q_1, q_2 )} ~,
\end{equation}
on the $\chi_+$ spinors, where
\begin{equation}
 \mathbb{A}^{( +, q_1, q_2 )} = -q_1 A^{-1} \Gamma_z \mathbb{B}^{( + )} + q_2 \mathcal{A}^{(+)} ,
\end{equation}
for some $q_1, q_2\in {\mathbb{R}}$. Observe that for $q_1, q_2\not=0$, the holonomy of $ \hat{\nabla}^{(+, q_1, q_2)}$ is not in $Spin(7)$.
 Next define the modified Dirac-like operator
\begin{equation}
\mathscr{D}^{(+)}\equiv \Gamma^i \hat{\nabla}^{(+, q_1, q_2)}_{i}= \Gamma^i \nabla_i + \Gamma^i \Psi^{(+)}_i + 7 \mathbb{A}^{(+, q_1, q_2)}~.
\label{mdirac}
\end{equation}
It is clear that if $\chi_+$ is a Killing spinor, ie satisfies (\ref{rkse3}), then it is a zero mode of $\mathscr{D}^{(+)}$. Here we that prove the converse. In particular, we shall show that there is
 a choice of $q_1, q_2$ such that
all the zero modes of $\mathscr{D}^{(+)}$ are Killing spinors. Thus we shall establish
\begin{equation}
\nabla_i^{(+)} \chi_\pm = 0, \qquad \mathcal{A}^{(+)} \chi_\pm = 0, \qquad  \mathbb{B}^{(+)} \chi_\pm = 0 \Longleftrightarrow \mathscr{D}^{(+)} \chi_+=0~.
\end{equation}
The proof relies on global properties of $M^7$, which we assume to be smooth, and compact without boundary.

To prove the theorem, let us assume that $\mathscr{D}^{(+)}\chi_+=0$ and consider the identity
\begin{equation}
 \nabla^2 \left\| \chi_+ \right\| ^2 = 2 \left\| \nabla \chi_+ \right\| ^2 + 2 \left\langle \chi_+, \nabla^2 \chi_+ \right\rangle~.
\end{equation}
The first term on the right hand side can be further rewritten in terms of the differential operator $\hat{\nabla}^{(+, q_1, q_2)}$ by completing the square as
\begin{eqnarray}
 2 \left\| \nabla \chi_+ \right\| ^2 &=& 2 \left\| \hat{\nabla}^{(+, q_1, q_2)} \chi_+ \right\| ^2 - 4 \left\langle \chi_+, \left( \Psi^{( + ) i \dagger} + \mathbb{A}^{(+, q_1, q_2) \dagger} \Gamma^i \right) \nabla_i \chi_+ \right\rangle
 \cr
 && \qquad - 2 \left\langle \chi_+, \left( \Psi^{( + ) i \dagger} + \mathbb{A}^{(+, q_1, q_2) \dagger} \Gamma^i \right) \left( \Psi^{( + )}_i + \Gamma_i \mathbb{A}^{(+, q_1, q_2)} \right) \chi_+ \right\rangle
 \cr
 &=& 2 \left\| \hat{\nabla}^{(+, q_1, q_2)} \chi_+ \right\| ^2 - 4 \left\langle \chi_+, \Psi^{( + ) i \dagger} \nabla_i \chi_+ \right\rangle
 \cr
 && \qquad - 2 \left\langle \chi_+, \left( \Psi^{( + ) i \dagger} - \mathbb{A}^{(+, q_1, q_2) \dagger} \Gamma^i \right) \left( \Psi^{( + )}_i + \Gamma_i \mathbb{A}^{(+, q_1, q_2)} \right) \chi_+ \right\rangle ,
\end{eqnarray}
while the second term can be rewritten using the identity $\slashed {\nabla}^2 = \nabla^2  - \frac{1}{4} R^{(7)}$, and  $\mathscr{D}^{(+)}\chi_+=0$, as
\begin{eqnarray}
 2 \left\langle \chi_+, \nabla^2 \chi_+ \right\rangle &=& 2 \left\langle \chi_+, \Gamma^i \nabla_i \left( \Gamma^j \nabla_j \chi_+ \right) \right\rangle + \frac{1}{2} R^{(7)} \left\| \chi_+ \right\| ^2
 \cr
 &=& \frac{1}{2} R^{(7)} \left\| \chi \right\| ^2 - 2 \left\langle \chi_+, \nabla_i \left( \Gamma^i \Gamma^j \Psi^{( + )}_j + 7 \Gamma^i \mathbb{A}^{(+, q_1, q_2)} \right) \chi_+ \right\rangle
 \cr
 &&\qquad - 2 \left\langle \chi_+, \left( \Gamma^i \Gamma^j \Psi^{( + )}_j + 7 \Gamma^i \mathbb{A}^{(+, q_1, q_2)} \right) \nabla_i \chi_+ \right\rangle .
\end{eqnarray}
Combining these, $\nabla^2 \left\| \chi_+ \right\| ^2$ can be rewritten as,
\begin{eqnarray}
 \nabla^2 \left\| \chi_+ \right\| ^2 &=& 2 \left\| \hat{\nabla}^{( +, q_1, q_2 )} \chi_+ \right\| ^2 + \frac{1}{2} R^{(7)} \left\| \chi_+ \right\| ^2
 \cr
 && \qquad + \left\langle \chi_+, \left[ -4 \Psi^{(+) i \dagger} - 2 \Gamma^i \Gamma^j \Psi^{(+)}_j - 14 q_1 A^{-1} \Gamma^{z i} \mathbb{B}^{(+)} - 14 q_2 \Gamma^i \mathcal{A}^{(+)} \right] \nabla_i \chi_+ \right\rangle
 \cr
 && \qquad + \left\langle \chi_+, -2 \left( \Psi^{(+) i \dagger} - \mathbb{A}^{(+, q_1, q_2) \dagger} \Gamma^i \right) \left( \Psi^{(+)}_i + \Gamma_i \mathbb{A}^{(+, q_1, q_2)}  \right) \chi_+ \right\rangle
 \cr
 && \qquad + \left\langle \chi_+, \nabla_i \left[ -2 \Gamma^i \Gamma^j \Psi^{(+)}_j - 14 q_1 A^{-1} \Gamma^{z i} \mathbb{B}^{(+)} - 14 q_2 \Gamma^i \mathcal{A}^{(+)} \right] \chi_+ \right\rangle \label{eq:laplacian_expansion}
\end{eqnarray}
where
\begin{eqnarray}
 \Psi^{(+) \dagger}_i = \frac{1}{8} \slashed{G}_i~,~~~
 \mathbb{B}^{(+) \dagger} =  -\frac{c_1 + c_2}{2 \ell}~,~~~
 \mathcal{A}^{(+) \dagger} = \slashed{\partial} \Phi + \frac{c_1}{\ell} A^{-1} \Gamma_z + \frac{1}{12} \slashed{G}~.
\end{eqnarray}

Of the terms on the right hand side of \eqref{eq:laplacian_expansion}, the first term is proportional to the gravitino Killing spinor equation squared, and so we expect that the remaining terms will be equal to some combination of the algebraic KSEs. The third term includes a derivative of $\chi_+$, however, and so we will attempt to write it in the form
\begin{equation}
 \alpha^i \nabla_i \left\| \chi_+ \right\| ^2 + \left\langle \chi_+, \mathcal{F} \Gamma^i \nabla_i \chi_+ \right\rangle = \alpha^i \nabla_i \left\| \chi_+ \right\| ^2 - \left\langle \chi_+, \mathcal{F} \left( \Gamma^i \Psi^{(+)}_i + 7 \mathbb{A}^{(+, q_1, q_2)} \right) \chi_+ \right\rangle~,
\end{equation}
for some vector $\alpha$ and Clifford algebra element  $\mathcal{F}$ that depend on the fields.
In terms of the fields, the third term in the right hand side of (\ref{eq:laplacian_expansion}) can be rewritten as
\begin{eqnarray}
 & \left\langle \chi_+, \left[ -4 \Psi^{(+) i \dagger} - 2 \Gamma^i \Gamma^j \Psi^{(+)}_{j} - 14 q_1 A^{-1} \Gamma^{z i} \mathbb{B}^{(+)} - 14 q_2 \Gamma^i \mathcal{A}^{(+)} \right] \nabla_{i} \chi_+ \right\rangle
 \\ \nonumber
 &= \left\langle \chi_+, \left[ \frac{7}{\ell} A^{-1} \Gamma^{z i} \left( q_1 c_1 + q_1 c_2 + 2 q_2 c_1 \right) - 14 q_2 \Gamma^i \slashed{\partial} \Phi \right. \right.
 \\ \nonumber
 & \qquad \qquad \left. \left. + \frac{1 + 14 q_2}{4} \slashed{G}^i + \frac{3 + 14 q_2}{12} \Gamma \slashed{G}^i \right] \nabla_i \chi_+ \right\rangle .
\end{eqnarray}
Thus, we find that it can be separated as outlined above if and only if $ q_2 = -\frac{1}{7} $. We will use this value of $q_2$ from here on. Then we  find that
\begin{eqnarray}
 & \left\langle \chi_+, \left[ -4 \Psi^{(+) i \dagger} - 2 \Gamma^i \Gamma^j \Psi^{(+)}_{j} - 14 q_1 A^{-1} \Gamma^{z i} \mathbb{B}^{(+)} + 2 \Gamma^i \mathcal{A} \right] \nabla_{i} \chi_+ \right\rangle
 \nonumber \\
 &= \left\langle \chi_+, \left[ \frac{1}{\ell} A^{-1} \Gamma^{z i} \left( 7 q_1 c_1 + 7 q_1 c_2 - 2 c_1 \right) + 2 \Gamma^i \slashed{\partial} \Phi \right. \right.
 \nonumber \\
 & \qquad \qquad \left. \left. - \frac{1}{4} \slashed{G}^i + \frac{1}{12} \Gamma \slashed{G}^i \right] \nabla_i \chi_+ \right\rangle ,
\end{eqnarray}
and so, factoring out a $\Gamma^i$ on the right,
\begin{equation}
 \mathcal{F} = \frac{1}{\ell} A^{-1} \Gamma_z \left( 7 q_1 c_1 + 7 q_1 c_2 - 2 c_1 \right) - 2 \slashed{\partial} \Phi - \frac{1}{12} \slashed{G}~,
\end{equation}
and $ \alpha_i = 2 \partial_i \Phi $.

The ${\cal{F}}$ term part of the third term of \eqref{eq:laplacian_expansion}  can be combined with the fourth term of \eqref{eq:laplacian_expansion} to give
\begin{eqnarray} \nonumber
 & \left\langle \chi_+, -2 \left( \Psi^{( + ) i \dagger} + q_1 A^{-1} \mathbb{B}^{(+) \dagger} \Gamma^{z i} + \frac{1}{7} \mathcal{A}^{(+) \dagger} \Gamma^i + \frac{1}{2} \mathcal{F} \Gamma^i \right) \right.
\nonumber \\
 & \qquad \left. \left( \Psi^{( + )}_i + q_1 A^{-1} \Gamma_{z i} \mathbb{B}^{(+)} - \frac{1}{7} \Gamma_i \mathcal{A}^{(+)} \right) \chi_+ \right\rangle
\nonumber  \\
 &= \left\langle \chi_+, -2 \left[ \frac{1}{\ell} \left( 3 q_1 c_1 + 3 q_1 c_2 - \frac{6 c_1}{7} \right) A^{-1} \Gamma^{z i} - \frac{6}{7} \slashed{\partial} \Phi \Gamma^i + \frac{1}{28} \slashed{G}^i + \frac{5}{168} \Gamma \slashed{G}^i \right] \right.
\nonumber  \\
 & \qquad \left.  \left[ -\frac{1}{\ell} \left( \frac{q_1 c_1 + q_1 c_2}{2} - \frac{c_1}{7} \right) A^{-1} \Gamma_{z i} - \frac{1}{7} \Gamma_i \slashed{\partial} \Phi - \frac{5}{56} \slashed{G}_i + \frac{1}{84} \Gamma \slashed{G}_i \right] \chi_+ \right\rangle
\nonumber \\
 &= \left\langle \chi_+, \left[ -\frac{3}{7 \ell^2} \left( 7 q_1 c_1 + 7 q_1 c_2 - 2 c_1 \right) ^2 A^{-2} - \frac{12}{7} (d\Phi)^2 - \frac{1}{21} \partial_i \Phi \Gamma \slashed{G}^i \right. \right.
\nonumber \\
 & \qquad \qquad \left. \left. - \frac{1}{42 \ell} \left( 7 q_1 c_2 + 7 q_1 c_1 - 2 c_1 \right) A^{-1} \slashed{G} \Gamma_z - \frac{1}{504} \slashed{G} \slashed{G} - \frac{1}{24} G^2 \right] \chi_+ \right\rangle \label{eq:algebraic_product} .
\end{eqnarray}

The last term on the right hand side of \eqref{eq:laplacian_expansion} is the only term involving derivatives of the fields other than $\Phi$ and the second derivative of $\Phi$. However, we can use the Bianchi identity and the $\Phi$ field equation to rewrite this term as,
\begin{eqnarray} \nonumber
 & \left\langle \chi_+, \nabla_i \left[ -2 \Gamma^i \Gamma^j \Psi^{( + )}_j - 2 A^{-1} \Gamma^{z i} \mathbb{B}^{(+)} + 2 \Gamma^i \mathcal{A} \right] \chi_+ \right\rangle
 \nonumber \\
 &= \left\langle \chi_+, \left[ 2 \nabla^2 \Phi + \frac{1}{48} \slashed{dG} \right] \chi_+ \right\rangle
 \nonumber \\
 &= \left\langle \chi_+, \left[ \frac{4}{\ell^2} A^{-2} + 4 (d\Phi)^2 - \frac{1}{6} G^2 \right] \chi_+ \right\rangle , \label{eq:field_eqns_and_bianchis}
\end{eqnarray}
and we can use the scalar part of the Einstein equation to rewrite the second term on the right hand side of \eqref{eq:laplacian_expansion} as
\begin{equation} \label{eq:curvature}
 \frac{1}{2} R^{(7)} \left\| \chi_+ \right\| ^2 = \left\langle \chi_+, \left[ -\frac{2}{\ell^2} A^{-2} - 2 (d\Phi)^2 + \frac{5}{24} G^2 \right] \chi_+ \right\rangle .
\end{equation}
Now we write the sum of \eqref{eq:algebraic_product}, \eqref{eq:field_eqns_and_bianchis}, and \eqref{eq:curvature}, as a linear combination of $ \left\| \mathbb{B}^{(+)} \chi_+ \right\| ^2 $, $\left\langle \Gamma_z \mathbb{B}^{(+)} \chi_+, \mathcal{A}^{(+)} \chi_+ \right\rangle$, and $\left\| \mathcal{A}^{(+)} \chi_+ \right\| ^2$.
In particular, the sum of \eqref{eq:algebraic_product}, \eqref{eq:field_eqns_and_bianchis}, and \eqref{eq:curvature} is given by
\begin{eqnarray}
 & \left\langle \chi_+, \left[  \frac{1}{\ell^2} \left( \frac{2}{7} - 42 {q_1}^2 + 12 q_1 + 12 q_1 c_1 c_2 - 42 {q_1}^2 c_1 c_2 \right) A^{-2} + \frac{2}{7} (d\Phi)^2 - \frac{1}{21} \partial_i \Phi \Gamma \slashed{G}^i \right. \right.
 \nonumber \\
 & \qquad \left. \left. - \frac{1}{42 \ell} \left( 7 q_1 c_2 + 7 q_1 c_1 - 2 c_1 \right) A^{-1} \slashed{G} \Gamma_z - \frac{1}{504} \slashed{G} \slashed{G} \right] \chi_+ \right\rangle ,
\end{eqnarray}
whereas
\begin{eqnarray}
 \left\| \mathbb{B}^{(+)} \chi_+ \right\| ^2 &=& \frac{1 + c_1 c_2}{2 \ell^2} \left\| \chi_+ \right\| ^2~,
 \cr
 \left\langle \Gamma_z \mathbb{B}^{(+)} \chi_+, \mathcal{A}^{(+)} \chi_+ \right\rangle
   &=& \left\langle \chi_+, \left[ -\frac{1}{2 \ell^2} \left( 1 + c_1 c_2 \right) A^{-1} - \frac{1}{24 \ell} \left( c_1 + c_2 \right) \slashed{G} \Gamma_z \right] \chi_+ \right\rangle~,
   \cr
 \left\| \mathcal{A}^{(+)} \chi_+ \right\| ^2
 &=& \langle \chi_+, [ (d\Phi)^2 + \frac{1}{\ell^2} A^{-2} - \frac{1}{6} \partial_i \Phi \Gamma \slashed{G}^i + \frac{c_1}{6 \ell} A^{-1} \slashed{G} \Gamma_z
 \cr
 \quad &&
 - \frac{1}{144} \slashed{G} \slashed{G} ] \chi_+ \rangle~.
\end{eqnarray}
It follows that
\begin{eqnarray}
 \nabla^2 \left\| \chi_+ \right\| ^2 &-& 2 \partial_i \Phi \nabla^i \left\| \chi_+ \right\| ^2
 = \left\|  \hat{\nabla}^{(+, q_1, q_2)} \chi_+ \right\| ^2 + 28 \left( q_1 - 3 {q_1}^2 \right) A^{-2} \left\| \mathbb{B}^{(+)} \chi_+ \right\| ^2
 \cr
 &+& 4 q_1 A^{-1} \left\langle \Gamma_z \mathbb{B}^{(+)} \chi_+, \mathcal{A} \chi_+ \right\rangle + \frac{2}{7} \left\| \mathcal{A} \chi_+ \right\| ^2 .
\end{eqnarray}
This expression is suitable to apply the Hopf maximum principle on the scalar  function $\| \chi_+ \|^2 $ on $M^7$ as for  $ 0 < q_1 < \frac{2}{7} $
the right hand side of this equation is positive definite.  Assuming that the conditions
required for the maximum principle on the fields and $M^7$ apply, e.g. the fields are smooth and $M^7$ is compact without boundary, the only solutions
to the above equation are that  $\left\| \chi_+ \right\| ^2$ is constant, and that,
\begin{equation}
 \nabla^{(\pm)} \chi_\pm = 0~, \qquad \mathcal{A}^{(\pm)} \chi_\pm = 0~, \qquad  \mathbb{B}^{(\pm)} \chi_\pm = 0~.
\end{equation}
Thus $\chi_+$ is a Killing spinor which establishes the theorem.

\newsection{AdS$_3$ backgrounds with $dH\not=0$ }

\subsection{ Bianchi identities, field equations and KSEs}

Let us first consider the modifications that occur in the Bianchi identity, field equations and KSEs of heterotic theory up to two loops in sigma model
perturbation theory\footnote{We use the conventions and normalization of the field equations, Bianchi identities  and KSEs  of  \cite{tsimpis}.}.
The anomaly cancelation mechanism requires the modification of the Bianchi identity for $H$ as
\begin{equation}
 dH = -\frac{\alpha'}{4} \left[ \text{tr} ( \tilde{R} \wedge \tilde{R} ) - \text{tr} (F \wedge F) \right]+{\cal O}(\alpha'^2)~,
 \label{anom}
\end{equation}
where $\tilde R$ is the curvature of a connection on the spacetime $M$ which will not be specified at this stage, $F$ is the  curvature of the gauge sector connection of
the heterotic theory and $\alpha'$ is the string tension which also has the role of the loop parameter. Thus $dH$ is expressed as  the difference of two Pontryagin forms, one is that of the tangent space of space-time and the other is that of the gauge sector bundle.    Furthermore, global anomaly cancelation
requires in addition that the form on the right-hand-side of the anomalous Binachi identity  represents the trivial cohomology class in $H^4(M)$.  This
statement is modified upon the addition of NS5-brane sources but this will not be considered here.

In addition to the modification of the Bianchi identity, the field equations also get modified.  In particular up to two loops in sigma model perturbation theory \cite{hulltown}, the dilaton, 2-form gauge potential, and gauge sector connection field equations read
\begin{eqnarray}
 \nabla^2 \Phi &=& 2 ( d \Phi )^2 - \frac{1}{12} H^2 + \frac{\alpha'}{16} \left[ \tilde{R}_{M N S T} \tilde{R}^{M N S T} - F_{M N a b} F^{M N a b} \right]+{\cal O}(\alpha'^2) ,
 \nonumber \\
 \nabla^R H_{MNR} &=&  2 \partial^R \Phi H_{MNR}+{\cal O}(\alpha'^2) ,
 \nonumber \\
 \nabla^N F_{MN} + [A^N, F_{MN}] &=&  2 \partial^N \Phi F_{MN} + \frac{1}{2} H_{MNQ} F^{NQ}+{\cal O}(\alpha')~,
 \label{dilalpha}
\end{eqnarray}
and the Einstein equation is
\begin{equation}
 R_{M N} = \frac{1}{4} H^2_{M N} - 2 \nabla_M \nabla_N \Phi - \frac{\alpha'}{4} \left[ \tilde{R}_{M L S T} \tilde{R}_N{}^{L S T} - F_{M L ab} F_N{}^{L ab} \right]+{\cal O}(\alpha'^2) \ .
 \label{ealpha}
\end{equation}

Furthermore, the KSEs \cite{bergshoeff} are
\begin{eqnarray}
 \nabla_M \epsilon - \frac{1}{8} \slashed{H}_M \epsilon &=& 0+{\cal O}(\alpha'^2)~,
 \cr
 \big(\slashed{\partial}\Phi-{1\over12} \slashed{H}\big)\epsilon&=&0+{\cal O}(\alpha'^2)~,
 \cr
 \slashed{F}\epsilon&=&0+{\cal O}(\alpha')~.
 \label{ksealpha2}
\end{eqnarray}
In particular observe that the KSEs have the same form up to two loops in sigma model perturbation theory as that at the zeroth order. It is not known
how these equations are modified at higher orders.  The gauge indices of $F$ have been suppressed.

Before we proceed with the investigation of AdS$_3$ backgrounds, let us specify $\tilde R$. In  perturbative heterotic theory, the choice of $\tilde R$
is renormalization scheme dependent. In other words, one can choose as $\tilde R$ the curvature of any connection on $M$. However in most applications $\tilde R$ is chosen
to be the curvature $\check R$ of the  $\check \nabla=\nabla - \frac{1}{2} H$ connection on the spacetime. It is known that this choice has some key advantages.
 In particular it is required for the cancelation
of world sheet supersymmetry anomaly \cite{wssa} and also for the consistency of the anomalous Bianchi identity with the modified Einstein equations
 for  supersymmetric backgrounds. This has been used in the calculations of \cite{howe40, tsimpis} and recently emphasized  \cite{martelli}. The property of $\check R$ which is used to establish these is
that $\check R$ satisfies   instanton-like conditions, i.e. it satisfies the same conditions, to zeroth order in $\alpha'$, as those implied on $F$ by the gaugino KSE. To see this, consider the identity
 \begin{equation}
\hat R_{MN, RS}=\check R_{RS, MN}-{1\over2} dH_{MNRS}~.
\label{chechhatr}
\end{equation}
The integrability condition of the gravitino KSE gives $\hat R_{MN, RS} \Gamma^{RS}\epsilon=0$. As the right-hand-side of the anomalous Bianchi identity is
 of order $\alpha'$, it follows from (\ref{chechhatr}) that,
to zeroth order in $\alpha'$, $\check R_{MN, RS} \Gamma^{MN}\epsilon=0$ or equivalently $\check {\slashed R}\epsilon=0$ after suppressing the $SO(9,1)$ gauge
 indices.  This is the same condition as that satisfied by the curvature of the gauge sector $F$ in (\ref{ksealpha2}).

To find solutions in the perturbative case, it is understood that the fields
and Killing spinors are expanded in $\alpha'$ schematically as
\begin{equation}
\label{taylorex}
g= g^0+\alpha' g^1+ {\cal O}(\alpha'^2)~,~\epsilon=\epsilon^0+\alpha' \epsilon^1+{\cal O}(\alpha'^2)~,
\end{equation}
and similarly for the 3-form field strength, gauge potential and dilaton. Then the field equations and KSEs are solved order by order in $\alpha'$ to find the correction to the zeroth order fields.

Next consider the case that the corrections to the heterotic theory are taken to be exact up to and including two loops. In such a case, $\alpha'$ is not an expansion
parameter. The anomalous Bianchi identity (\ref{anom}), field equations, (\ref{dilalpha}) (\ref{ealpha}), and KSEs (\ref{ksealpha2}) do not receive
 further corrections from the ones that have been explicitly stated. However consistency of the anomalous Bianchi identity with the field equations requires that $\tilde R$ satisfies the same conditions as those implied
by the KSEs on the curvature $F$ of the gauge connection, ie $\tilde{\slashed  R}\epsilon=0$ after suppressing the gauge indices.
It is not apparent that such a connection always exists but there are existence theorems in many cases of interest.
Notice also the difference from the perturbation  theory as  $\tilde R$ cannot be identified with $\check R$.  This is because $dH$ does not vanish in the right-hand-side of (\ref{chechhatr}).

\subsection{AdS$_3$ backgrounds in perturbation theory}

Suppose that the symmetries of AdS$_3$ remain symmetries of the background after the $\alpha'$ corrections are taken into account.
In such a case, the fields up to two loops in perturbation theory will decompose as
\begin{eqnarray}
 ds^2 &=& 2 \mathbf{e}^+ \mathbf{e}^- + A^2 dz^2 + ds^2 \left( M^7 \right)+{\cal O}(\alpha'^2)~,
 \cr
  H &=& A X e^+ \wedge e^- \wedge dz + G +{\cal O}(\alpha'^2)~.
  \label{falpha}
 \end{eqnarray}
 This assumption is justified later.
 Furthermore, the field equations (\ref{dilalpha}) and (\ref{ealpha}) read
 \begin{eqnarray}
 \nabla^2 \Phi &=& -3 A^{-1} \partial_i A \partial^i \Phi + 2 (d\Phi)^2 - \frac{1}{12} G^2 + \frac{1}{2} X^2
 \cr
 && \qquad + \frac{\alpha'}{16} \left[ \check{R}_{i j, k \ell} \check{R}^{i j, k \ell} - F_{i j a b} F^{i j a b} \right]+{\cal{O}}(\alpha'^2)~,
 \cr
 \nabla^k G_{i j k} &=& -3 A^{-1} \partial^k A G_{i j k} + 2 \partial^k \Phi G_{i j k}+{\cal{O}}(\alpha'^2)~,
\end{eqnarray}
and the AdS component of the Einstein equation is unchanged,
\begin{equation}
 \nabla^2 \ln A = -\frac{2}{\ell^2} A^{-2} - 3 A^{-2} (dA)^2 + 2 A^{-1} \partial_i A \partial^i \Phi + \frac{1}{2} X^2+{\cal{O}}(\alpha'^2)~,
\end{equation}
while component on $M^7$ is now,
\begin{eqnarray}
 R^{(7)}_{i j} &= 3 \nabla_i \nabla_j \ln A + 3 A^{-2} \partial_i A \partial_j A + \frac{1}{4} G_{i k_1 k_2} G_j{}^{k_1 k_2} - 2 \nabla_i \nabla_j \Phi
 \nonumber \\
 & \qquad - \frac{\alpha'}{4} \left[ \check{R}_{i k, s t} \check{R}_j{}^{k, s t} - F_{i k a b} F_j{}^{k a b} \right] +{\cal{O}}(\alpha'^2),
\end{eqnarray}
where $i,j,k,\ell=1,2,\dots,7$  and we have assumed that $\check R$ and $F$ do not have components along the AdS$_3$ directions. As we shall see, this will follow
from the KSEs.

 In addition, one finds that
\begin{eqnarray}
 R^{(7)} &=& 3 \nabla^2 \ln A + 3 A^{-2} (dA)^2 + \frac{1}{4} G^2 - 2 \nabla^2 \Phi
 \cr
 && \qquad- \frac{\alpha'}{4} \left[ \check{R}_{i j, k \ell} \check{R}^{i j, k \ell} - F_{i j a b} F^{i j a b} \right]+{\cal{O}}(\alpha'^2)
 \cr
 &=& -\frac{6}{\ell^2} A^{-2} - 6 A^{-2} (dA)^2 + \frac{5}{12} G^2 + \frac{1}{2} X^2 + 12 A^{-1} \partial_i A \partial^i \Phi - 4 (d\Phi)^2
 \cr
 && \qquad - \frac{3 \alpha'}{8} \left[ \check{R}_{i j, k \ell} \check{R}^{ i j, k \ell} - F_{i j a b} F^{i j a b} \right]+{\cal{O}}(\alpha'^2) \ .
\end{eqnarray}
Similarly, the anomalous
 Bianchi identity of $H$ reads
\begin{eqnarray}
 dG &= -\frac{\alpha'}{4} \left[ \text{tr} \left( \check{R} \wedge \check{R} \right) - \text{tr} (F \wedge F) \right] +{\cal{O}}(\alpha'^2)~.
\end{eqnarray}
As we shall see imposing the requirement that spacetime supersymmetry is preserved by the higher order corrections simplifies
the above equations further.

\subsection{Geometry of $M^7$ for backgrounds with two supersymmetries}
\label{g2alpha}

In the perturbative approach to the heterotic string, one of the questions that arises is whether the higher order corrections preserve the spacetime supersymmetry
of the zeroth order background. In other words, whether there is a renormalization
scheme which preserves the spacetime supersymmetry order by order in perturbation theory. Here we shall not investigate the existence of such a scheme. Instead, we
shall derive the conditions for such a scheme to exist.

 We have shown that for AdS$_3$ backgrounds admitting two spacetime supersymmetries at zero order in $\alpha'$, $M^7$ has a $G_2$ structure compatible with a
 connection with skew-symmetric torsion. In particular at this order $dH=0$, and $A$ and $X$ are constant and $c_1=\frac{\ell}{2}AX=\pm 1$.
 The geometry of $M^7$ at this order is
described in section \ref{g2geom}.

The contribution in the terms proportional to $\alpha'$ in the field equations, Bianchi identities and KSEs comes from the fields at zeroth order in $\alpha'$.
These depend on $\check R$ and $F$.
At zeroth order, the spacetime factorizes into a product $AdS_3\times M^7$.  Furthermore
 the choice of torsion on $AdS_3$ is such that $\hat\nabla\vert_{AdS_3}$ and $\check \nabla\vert_{AdS_3}$ are either the left or right invariant parallelizing connection; $AdS_3$ is a group manifold. In either case,
 $\hat R\vert_{AdS_3}=\check R\vert_{AdS_3}=0$.  Therefore the contribution in the $\alpha'$ terms of field equations, Bianchi identities and KSEs comes only from the $\check R^{(7)}$ curvature of $M^7$.   Furthermore, the KSEs imply  that the gauge curvature $F$
 does not have components along AdS$_3$ and  is invariant under the isometries of AdS$_3$ up to gauge transformations. As a result all gauge
 invariant tensors constructed from $F$ are tensors on $M^7$ which do not depend on the coordinates of AdS$_3$. These justify the choice of $\check R$ and $F$
  made in the previous section.

 As the form of the gravitino KSE remains the same up to order $\alpha'^2$, this  implies that $A$ and $X$ are constant up to that order and that again $c_1=\frac{\ell}{2}AX=\pm 1$.
 Furthermore the metric and torsion of
 $AdS_3$ does not receive corrections at one loop, the form of the fields remains as in (\ref{falpha})  up to order ${\cal O}(\alpha'^2)$. The background
 remains factorized as $AdS_3\times M^7$ up to that order as well. Imposing all the above conditions on the fields, one finds that the anomalous Bianchi identity
 and field equations are simplified as in appendix B.

Next focusing on the geometry of $M^7$,  $M^7$  admits a $G_2$ structure compatible with a connection $\hat\nabla$ with skew-symmetric torsion $G$. As a consequence of the gravitino and dilatino KSEs,
$G$ is as given in (\ref{gg2flux}) up to order $\alpha'^2$.  Moreover all the KSEs and field equations are satisfied provided that
\begin{eqnarray}
d\big(e^{2\Phi} \star_7 \varphi\big)=0+O(\alpha'^2)~,~~~dG= -\frac{\alpha'}{4} \left[ \text{tr} \left( \check{R}^{(7)} \wedge \check{R}^{(7)} \right) - \text{tr} (F \wedge F) \right] +{\cal{O}}(\alpha'^2)~.
\label{g2conalpha}
\end{eqnarray}
The first condition is required for the existence of a connection with skew-symmetric torsion which is compatible
with the $G_2$ structure on $M^7$ while the second condition arises from the anomalous Bianchi identity. We have also assumed as in the $dH=0$ case  that all solutions $\epsilon$
 of the gravitino KSEs are also solutions of the gaugino KSE, $\slashed{F}\epsilon=0$. In this case, this implies that $F$ is a $G_2$ instanton on $M^7$,
and so it satisfies
\begin{eqnarray}
F_{ij}={1\over2}\star_7\varphi_{ij}{}^{km} F_{km}+{\cal{O}}(\alpha')~,
\label{g2fcon}
\end{eqnarray}
where we have suppressed the gauge indices.
This summarizes the geometry of $M^7$ up to order $\alpha'^2$.

\subsection{Extended supersymmetry}

Next let us investigate the geometry of AdS$_3$ backgrounds preserving 4, 6 and 8 supersymmetries up to order $\alpha'^2$.  The geometry of the associated zeroth order
backgrounds for which  $dH=0$ has already been described in section \ref{extendsusy}.

\subsubsection{Four supersymmetries}

These backgrounds are a special case of those we have described in the previous section that preserve two supersymmetries.  As a result up to
order $\alpha'^2$, the geometry is a product $AdS_3\times M^7$. The presence of two more supersymmetries restricts further the geometry of $M^7$.
As the form  of the gravitino and dilatino KSEs remain the same as that of the zeroth order fields, the geometric restrictions on the geometry of $M^7$ are similar to those in section \ref{extendsusy}.
The only difference here is that $dH\not=0$. In particular,   $M^7$ has an $SU(3)$ structure compatible with a connection with skew-symmetric torsion. So it admits a Killing vector field $\xi$ such that
\begin{eqnarray}
i_\xi G=k^{-1} dw+O(\alpha'^2)~,~~~i_\xi F=0+{\cal{O}}(\alpha')~,
\label{xigfalpha}
\end{eqnarray}
where $w(\xi)=k$.
Moreover, $M^7$
can be locally described
as a circle fibration of  a conformally balanced, $\theta_\omega=2d\Phi$, KT manifold $B^6$ with Hermitian form $\omega$ whose canonical bundle admits a connection $k^{-1} w$,   such that
\begin{eqnarray}
dw^{(2,0)}=0+O(\alpha'^2)~,~~~dw_{ij} \omega^{ij}=-2k^2+{\cal{O}}(\alpha'^2)~,
\label{wsu3alpha}
\end{eqnarray}
i.e. the canonical bundle is holomorphic and the connection satisfies the Hermitian-Einstein instanton condition, and in addition
\begin{eqnarray}
\hat\rho_{(6)}=dw+O(\alpha'^2)~,~k^{-2} dw\wedge dw+d G_{(6)}=-\frac{\alpha'}{4} \left[ \text{tr} \left( \check{R}^{(7)} \wedge \check{R}^{(7)} \right) - \text{tr} (F \wedge F) \right] +{\cal{O}}(\alpha'^2)~,
\nonumber \\
\label{eqnsu3alpha}
\end{eqnarray}
where
\begin{eqnarray}
\hat\rho^{(6)}_{ij}={1\over2} \hat R^{(6)}{}_{ij}{}^k{}_m I^m{}_k+{\cal{O}}(\alpha'^2)~,
\label{hatrho}
\end{eqnarray}
 is the curvature of the canonical bundle induced from the connection with torsion $G_{(6)}=-i_I d\omega$ on $B^6$ and $I$ is the complex structure of $B^6$.  The first condition is required for $M^7$ to admit an $SU(3)$ structure compatible with the connection with skew-symmetric torsion $G$ and
the second condition is required by the anomalous Bianchi identity (\ref{g2conalpha}), where now $i,j,k,m=1,2,\dots,6$. It is understood that the expression in the right-hand-side of the second equation in (\ref{eqnsu3alpha})
is evaluated at the zeroth order fields. The metric and torsion on $M^7$ are given from those of $B^6$ as in (\ref{6from7})
but now of course the fields on $B^6$ obey the equations (\ref{eqnsu3alpha}) above.

\subsubsection{Six supersymmetries}

The presence of additional supersymmetries restricts the geometry of $M^7$ further. In particular, the spacetime is still a product $AdS_3\times M^7$ up to order $\alpha'^2$.
The geometry of the zeroth order configuration has already been described in section \ref{extendsusy} and so $M^7$ is locally a $S^3$ fibration over a 4-dimensional manifold $B^4$.
As the gravitino and dilatino KSEs have the same form up to order $\alpha'^2$ as the zeroth order equations, it is expected that $M^7$ admits three $\hat\nabla$-parallel vector bilinears $\xi_{r'}$, $r'=1,2,3$. Thus $\xi_{r'}$ are isometries of the metric on $M^7$ and $i_{\xi_{r'}} H= k^{-1} dw_{r'}$ up to order $\alpha'^2$, where $w_{r'}(\xi_{s'})=k \delta_{r's'}$. As the geometry
of the spacetime is a product up to $\alpha'^2$, these commute with the isometries of AdS$_3$.
However, the gravitino and dilatino KSEs do not determine the Lie bracket algebra of $\xi_{r'}$'s.

To determine $[\xi_{r'}, \xi_{s'}]$, first note that the commutator of two isometries is an isometry. Then using $\hat\nabla \xi_{r'}=0$, we can establish the identities
\begin{eqnarray}
k^{-1}w_{[\xi_{r'}, \xi_{s'}]}=i_{\xi_{r'}} i_{\xi_{s'}} H~,~~~i_{[\xi_{r'}, \xi_{s'}]} H= k^{-1} dw_{[\xi_{r'}, \xi_{s'}]}+i_{\xi_{r'}} i_{\xi_{s'}} dH~.
\label{syriza}
\end{eqnarray}
Next note that $i_{\xi_{r'}} i_{\xi_{s'}} dH=0+{\cal O}(\alpha'^2)$.  This follows from the fact that both $\check R$ and $F$ contribute in $dH$ via the zeroth order fields
and so as a consequence of the gravitino and gaugino KSEs, $i_{\xi_{r'}} \check R=i_{\xi_{r'}} F=0$. In fact $F$ has to be a anti-self-dual instanton in the directions
transverse to AdS$_3$ and  $\xi_{r'}$.
As a consequence, the commutator $[\xi_{r'}, \xi_{s'}]$ is $\hat\nabla$-parallel up to order $\alpha'^2$. If $[\xi_{r'}, \xi_{s'}]$ is not expressed
in terms of $\xi_{r'}$, the holonomy of $\hat\nabla$ is reduced to $\{1\}$ implying that the zeroth order backgrounds are group manifolds. Such backgrounds preserve
8 supersymmetries and will be investigated below. Thus $[\xi_{r'}, \xi_{s'}]$ must close on $\xi_{t'}$.
Furthermore, one can use the Bianchi identity
\begin{eqnarray}
\hat R_{M[N,PQ]}=-{1\over3} \hat\nabla_M H_{NPQ}+{1\over6} dH_{MNPQ}~,
\label{bxxx}
\end{eqnarray}
to show that $dw^{r'}$ restricted on the directions transverse to AdS$_3$ and $\xi_{r'}$ is $\hat\nabla$-parallel.  Then an analysis similar to that we have done for
heterotic horizons \cite{hethor} reveals that $\xi_{r'}$ close to a $\mathfrak{su}(2)$ algebra.  As a result, $M^7$ is locally a $S^3$ fibration over a 4-dimensional manifold $B^4$.
The geometry can be described exactly as in the zeroth order case but the various formulae are now valid up to order $\alpha'^2$.
The only modification occurs in the equation for the dilaton which now reads
\begin{eqnarray}
\mathring{\nabla}^2 e^{2\Phi}=-{1\over2} ({\cal F}^{\mathrm{ad}})^2+{3\over8} k^2 e^{4\Phi}+{\alpha'\over8} \big(\tilde R^{(4)}{}^2- F^2\big)+{\cal O}(\alpha'^2)~,
\label{fcon6alpha}
\end{eqnarray}
where the inner products are taken with respect to the $d\mathring s^2$ metric.
The additional $\alpha'$ contribution is due to the anomalous Bianchi identity of $H$.

\subsubsection{Eight supersymmetries}

The backgrounds with 8 supersymmeries can be investigated in a way similar to those with  6 supersymmetries described in the previous  section.
However there are some differences.  As we have already mentioned at zeroth order in $\alpha'$,  section \ref{extendsusy},  $B^4$  is a hyper-K\"ahler manifold and ${\cal F}^{\mathrm{sd}}=0$. Up to order $\alpha'^2$, the spacetime remains a product $AdS_3\times M^7$. The investigation of the closure properties of
the three $\hat\nabla$-parallel vector field $\xi_{r'}$ on $M^7$ is not necessary.  This is because it is a consequence of the gravitino and dilatino
KSEs that these vector fields close to a $\mathfrak{su}(2)$ algebra \cite{hetclas}. The metric and torsion are given as in (\ref{mtor6}) but now the formulae
are valid up to order $\alpha'^2$.  The only modification from the zeroth order equations is that the dilaton equation now reads
\begin{eqnarray}
\mathring{\nabla}^2 e^{2\Phi}=-{1\over2} ({\cal F}^{\mathrm{ad}})^2+
{\alpha'\over8} \big(\tilde R^{(4)}{}^2- F^2\big)+{\cal O}(\alpha'^2)~,
\label{fcon8alpha}
\end{eqnarray}
where the metric $d\mathring{s}(B^4)$ is the zeroth order hyper-K\"ahler metric and the inner products have been taken with respect to it.

For compact $B^4$, at zeroth order ${\cal F}^{\mathrm{ad}}=0$, and in this case $M^7= S^3\times B^4$ up to discrete identifications.
As a consequence, the worldsheet action of the string factorizes into a sum of a WZW model on $S^3$ and a sigma model on the
hyper-K\"ahler manifold $B^4$.  The latter has (4,0) worldsheet supersymmetry and as a result is ultraviolet finite \cite{howegp}.  However,
in the presence of an anomaly, the couplings are corrected order by order in $\alpha'$ as a consequence of maintaining manifest
(4,0) supersymmetry in perturbation theory \cite{howe40}.

\subsection{Truncation  to two loops}

Suppose now that the theory up to two loops is exact. In such a case, the geometry of the solutions  has to be re-examined as several arguments
that have been applied in previous cases have been based on the closure of $H$ either to all orders or at the zeroth
order in perturbation theory. Moreover $\alpha'$ has been treated as an arbitrary parameter. None of these
two assumptions are valid any longer. Nevertheless, there is a simplifying assumption. This is that the backgrounds have the
symmetries of AdS$_3$.  In particular, the fields can be written as (\ref{fansatz}). The KSEs are
\begin{eqnarray}
 \nabla_M \epsilon - \frac{1}{8} \slashed{H}_M \epsilon = 0~,~~~
 \big(\slashed{\partial}\Phi-{1\over12} \slashed{H}\big)\epsilon=0~,~~~\slashed{F}\epsilon=0~.
 \label{kseexact}
\end{eqnarray}
We  also assume that the gaugino KSE has as many Killing spinors
as the gravitino KSE.

\subsubsection{Two supersymmetries}

The $G_2$ case is rather straightforward. As the form of the gravitino and dilatino KSEs in (\ref{kseexact}) is the same as that for $dH=0$ backgrounds and the fields are invariant
under the symmetries of AdS$_3$, one finds that the gravitino KSE
 implies that $A,X$ are constant and $c_1=\frac{\ell}{2}AX=\pm 1$. As a result, the geometry locally decomposes as $AdS_3\times M^7$.  The geometry of $M^7$ can now be described as in the perturbative
case with the only difference that now the equations are exact.  In particular, $M^7$ admits a $G_2$ structure compatible with a connection with skew-symmetric torsion.
This $G_2$ structure is further restricted by the KSEs, Bianchi identities and field equations as
\begin{eqnarray}
d\big(e^{2\Phi} \star_7 \varphi\big)=0~,~~~dG= -\frac{\alpha'}{4} \left[ \text{tr} \left( \tilde{R}^{(7)} \wedge \tilde{R}^{(7)} \right) - \text{tr} (F \wedge F) \right] ~,
\label{g2cconexact}
\end{eqnarray}
where $\varphi$ is the fundamental $G_2$ 3-form, $G=k\varphi+ e^{2\Phi} \star_7 d(e^{-2\Phi} \varphi)$, and $\tilde R$ and $F$ are $G_2$ instantons, ie
\begin{eqnarray}
\tilde R^{(7)}_{ij,pq}= {1\over2}\star_7\varphi_{ij}{}^{km} \tilde R^{(7)}_{km, pq}~,~~~F_{ij}={1\over2}\star_7\varphi_{ij}{}^{km} F_{km}~.
\label{g2fconexact}
\end{eqnarray}
The condition on $F$ follows from the gaugino KSE. Observe that $\check R$, which is no longer a $G_2$ instanton because of (\ref{chechhatr}) and $dH\not=0$, has now been replaced with $\tilde R^{(7)}$.
Moreover $\alpha'$ in (\ref{g2cconexact}) is a constant rather than a parameter.

\subsubsection{Four supersymmetries}

The geometry of these backgrounds also factorizes as $AdS_3\times M^7$. Moreover, $M^7$ admits a $SU(3)$ structure compatible with a connection $\hat\nabla$ with skew-symmetric torsion.
There are 4 vector spinor bilinears and there is a basis such that 3 of them generate an $\mathfrak{sl}(2,{\mathbb{R}})$   symmetry of AdS$_3$.  As these 4 vector bilinears are $\hat\nabla$-parallel,
their commutator is $
[\xi_a, \xi_b]=i_{\xi_a} i_{\xi_b} H$.
Since the geometry factorizes as $AdS_3\times M^7$, it turns out that the commutator of the generators of $\mathfrak{sl}(2,{\mathbb{R}})$ with the fourth vector bilinear
vanishes, and so the symmetry algebra of the spacetime is $\mathfrak{sl}(2,{\mathbb{R}})\oplus \mathfrak{u}(1)$.

The rest of the analysis is similar to that we have described for the perturbative case. In particular, the equations (\ref{xigfalpha}), (\ref{wsu3alpha}),
(\ref{eqnsu3alpha}) and (\ref{hatrho}) are still valid but now exactly. The only modification is in the second equation in (\ref{eqnsu3alpha}) which now reads
\begin{eqnarray}
k^{-2} dw\wedge dw+d G_{(6)}=-\frac{\alpha'}{4} \left[ \text{tr} \left( \tilde{R}^{(6)} \wedge \tilde{R}^{(6)} \right) - \text{tr} (F \wedge F) \right]~,
\end{eqnarray}
where $\tilde{R}^{(6)}$ is a $\mathfrak{su}(3)$ instanton on $B^6$, ie $\tilde{R}^{(6)}$ is a (1,1)-form and $\omega$-traceless. This condition is also satisfied by $F$ because of the gaugino KSE.

\subsubsection{Six supersymmetries}

The geometry factorizes as $AdS_3\times M^7$ and $M^7$ admits an $SU(2)$ structure compatible with a connection with skew-symmetric torsion $\hat\nabla$. The spacetime admits 6 vector Killing spinor bilinears. Three of these span an $\mathfrak{sl}(2,{\mathbb{R}})$  symmetry of AdS$_3$,
and the other three $\xi_{r'}$ are $\hat\nabla$-parallel on $M^7$ and commute with those generating the $\mathfrak{sl}(2,{\mathbb{R}})$.
We shall argue that for non-trivial backgrounds the commutator of these three vector field must close in the set.
To see this,   consider the identities in (\ref{syriza}). As $\xi_{r'}$ are Killing, their commutator is also Killing.
Furthermore, the term $i_{\xi_{r'}} i_{\xi_{s'}} dH$ in the second equation in (\ref{syriza}) vanishes.  This is because
we have assumed that the connections that contribute  in the anomalous Bianchi identity are those that satisfy the gaugino KSE. For all these $i_{\xi_{r'}} F=i_{\xi_{r'}} \tilde R=0$.
As a result, if $\xi_{r'}$ and $\xi_{s'}$ are $\hat\nabla$-parallel, so is the commutator $[\xi_{r'}, \xi_{s'}]$. If the commutator
does not close in the set $\xi_{r'}$, the holonomy of $\hat\nabla$ will reduce to $\{1\}$. As a result the curvature of $\hat\nabla$ vanishes.  If this is the case,  the contribution
to the anomalous Bianchi identity must vanish as well as the connections that contribute to it have zero curvature. This is implied by our assumption that all solutions to
the gravitino KSE are also solutions of the gaugino one.
 For such backgrounds backgrounds $dH=0$ and so the spacetime  is a group manifold which preserves 8 supersymmetries.  Thus for  backgrounds with strictly six supersymmetries, we shall take
 that $[\xi_{r'}, \xi_{s'}]$ closes in the set $\xi_{r'}$. Then it can be shown using (\ref{bxxx}) that the symmetry group of the spacetime generated by the vector spinor bilinears is $\mathfrak{sl}(2,{\mathbb{R}})\oplus \mathfrak{su}(2)$.

The rest of the investigation of the geometry is similar to that we have done in the perturbative case. The only difference is that now
\begin{eqnarray}
\mathring{\nabla}^2 e^{2\Phi}=-{1\over2} ({\cal F}^{\mathrm{ad}})^2+{3\over8} k^2 e^{4\Phi}+{\alpha'\over8} \big(\tilde R^{(4)}{}^2- F^2\big)~,
\label{fcon6exact}
\end{eqnarray}
where $\tilde R^{(4)}$ and $F$ are anti-self-dual instantons on $B^4$ and  the inner products are taken with respect to the $d\mathring s^2$ metric. $B^4$ is a 4-manifold with vanishing self-dual Weyl tensor and metric $d\mathring{s}(B^4)$.

\subsubsection{Eight supersymmetries}

The investigation of the geometry of these backgrounds is simpler than that described in the previous section for backgrounds preserving 6 supersymmetries.
First the geometry factorizes as $AdS_3\times M^7$ and $M^7$ admits a connection with skew-symmetric torsion compatible with a $SU(2)$ structure.
As in the previous case, $M^7$ admits 3 $\hat\nabla$-parallel Killing spinor bilinears $\xi_{r'}$ which commute with another three which span an $\mathfrak{sl}(2,{\mathbb{R}})$
symmetry of AdS$_3$. Furthermore the gravitino and dilatino KSEs imply that the symmetry algebra  of these backgrounds is $\mathfrak{sl}(2,{\mathbb{R}})\oplus \mathfrak{su}(2)$.
The analysis of the geometry proceeds as in the perturbative case. In particular, $M^7$ is an $S^3$ fibration over a hyper-K\"ahler manifold $B^4$ with metric $d\mathring{s}(B^4)$.
The only difference from the perturbative case is that now
\begin{eqnarray}
\mathring{\nabla}^2 e^{2\Phi}=-{1\over2} ({\cal F}^{\mathrm{ad}})^2+
{\alpha'\over8} \big(\tilde R^{(4)}{}^2- F^2\big)~,
\label{fcon8exact}
\end{eqnarray}
where $\tilde R^{(4)}$ and $F$ are anti-self-dual instantons on $B^4$.

\subsection{Lichnerowicz type Theorem on $ \sigma_+ $, $ \tau_+ $}

The Lichnerowicz type theorem has to be re-examined in the presence of $\alpha'$ corrections and in the case that the theory is truncated to two loops.
Again, we shall focus on $M^7$, and
define the modified Dirac-like operator as in (\ref{mdirac}) but now $dG\not=0$. Furthermore, we assume the Bianchi identities and field equations
of appendix B but now we shall include the $\alpha'$ terms,
replacing the $\check R^{(7)}$ terms with ${\tilde{R}}$, and replacing
$F$ with ${\tilde{F}}$ where ${\tilde{R}}$ and ${\tilde{F}}$ are
arbitrary curvatures of $TM^7$ and the gauge sector bundle
respectively.
In particular
$\tilde R$ and $\tilde F$ are not restricted by the KSEs. For the truncated theory at two loops, we take the equations in appendix B as exact but again with
$\check R^{(7)}$ and $F$ replaced with $\tilde R$ and $\tilde F$.

The derivation of \eqref{eq:algebraic_product} is unaffected, but \eqref{eq:field_eqns_and_bianchis} becomes
\begin{eqnarray} \nonumber
 & \left\langle \chi_+, \nabla_i \left[ -2 \Gamma^i \Gamma^j \Psi^{( + )}_j - 2 A^{-1} \Gamma^{z i} \mathbb{B}^{(+)} + 2 \Gamma^i \mathcal{A}^{(+)} \right] \chi_+ \right\rangle
\nonumber  \\
 &= \left\langle \chi_+, \left[ 2 \nabla^2 \Phi + \frac{1}{48} \slashed{dG} \right] \chi_+ \right\rangle
 \nonumber \\
 &= \left\langle \chi_+, \left[ \frac{4}{\ell^2} A^{-2} + 4 (d\Phi)^2 - \frac{1}{6} G^2 + \frac{\alpha'}{8} \left[ \tilde{R}_{i j, k \ell} \tilde {R}^{ i j, k \ell} -\tilde F_{i j a b}\tilde F^{i j a b} \right] \right. \right.
\nonumber  \\
 & \qquad \left. \left. + \frac{\alpha'}{32} \left[ \tilde {R}_{i_1 i_2, j k} \tilde{R}_{i_3 i_4,}{}^{j k} - \tilde F_{i_1 i_2 a b} \tilde  F_{i_3 i_4}{}^{a b} \right] \Gamma^{i_1 i_2 i_3 i_4} \right] \chi_+ \right\rangle ,
\end{eqnarray}
and \eqref{eq:curvature} also picks up an $\alpha'$ term;
\begin{eqnarray}
 \frac{1}{2} R^{(7)} \left\| \chi_+ \right\| ^2 &= \left\langle \chi_+, \left[ -\frac{2}{\ell^2} A^{-2} - 2 (d\Phi)^2 + \frac{5}{24} G^2 \right. \right.
 \nonumber \\
 & \qquad \left. \left. - \frac{3 \alpha'}{16} \left[ \tilde{R}_{i j, k \ell} \tilde{R}^{ i j, k \ell} - \tilde F_{i j a b} \tilde F^{i j a b} \right] \right] \chi_+ \right\rangle .
\end{eqnarray}
On combining these expressions we obtain
\begin{eqnarray}
\label{lichcorr}
 \nabla^2 \left\| \chi_+ \right\| ^2 - 2 \partial_i \Phi \nabla^i \left\| \chi_+ \right\| ^2
 &=& \left\|  \hat{\nabla}^{(+, q_1, q_2)} \chi_+ \right\| ^2 + 28 \left( q_1 - 3 {q_1}^2 \right) A^{-2} \left\| \mathbb{B}^{(+)} \chi_+ \right\| ^2
 \nonumber \\
 &+& 4 q_1 A^{-1} \left\langle \Gamma_z \mathbb{B}^{(+)} \chi_+, \mathcal{A} \chi_+ \right\rangle + \frac{2}{7} \left\| \mathcal{A} \chi_+ \right\| ^2
+ {\alpha' \over 32} \parallel {\slashed{\tilde{F}}} \chi_+ \parallel^2
\nonumber \\
&-&{\alpha' \over 32} \langle {\tilde{R}}_{\ell_1 \ell_2,mn} \Gamma^{\ell_1 \ell_2} \chi_+, {\tilde{R}}_{p_1 p_2,}{}^{mn} \Gamma^{p_1 p_2} \chi_+ \rangle \ ,
\end{eqnarray}
where we have suppressed the gauge index contraction in the
$\parallel {\slashed{\tilde{F}}} \chi_+ \parallel^2$ term, and $q_2=-{1 \over 7}$.

We shall first consider the case of perturbation theory, and set
${\tilde{R}}=\check R^{(7)}$. We begin by systematically analysing the conditions imposed
by ({\ref{lichcorr}}) order by order in $\alpha'$. {\footnote{We remark that
in perturbation theory, the RHS of ({\ref{lichcorr}}) is explicitly determined only up to first order in
$\alpha'$. The $\alpha'^2$ terms are not known, as one would
require the corresponding $\alpha'^2$ corrections to the
Dirac operator, as well as $dG$ and $R^{(7)}$ and $\nabla^2 \Phi$, in order to fix the $\alpha'^2$ terms.}}

To zeroth order in $\alpha'$, one obtains (provided that
$0<q_1<{2 \over 7}$), the conditions
\begin{eqnarray}
\label{cond1}
\nabla^{(+)} \chi_+ = 0+{\cal O}(\alpha')~, \qquad \mathcal{A}^{(+)} \chi_+ = 0+{\cal O}(\alpha')~, \qquad  \mathbb{B}^{(+)} \chi_+ = 0+{\cal O}(\alpha')~.
\end{eqnarray}
The condition $\nabla^{(+)} \chi_+ = 0+{\cal O}(\alpha')$ implies the integrability condition
\begin{eqnarray}
{\hat{R}}^{(7)}_{mn, \ell_1 \ell_2} \Gamma^{\ell_1 \ell_2} \chi_+ =  0+{\cal O}(\alpha') \ .
\end{eqnarray}
This in turn implies that
\begin{eqnarray}
{\check{R}}^{(7)}_{\ell_1 \ell_2,mn} \Gamma^{\ell_1 \ell_2} \chi_+ =  0+{\cal O}(\alpha') \ .
\end{eqnarray}
It follows that the final term in ({\ref{lichcorr}}) is in fact at least
of order $\alpha'{}^3$, and so can be ignored.

It remains to show that ({\ref{lichcorr}}) implies the KSEs to linear order in $\alpha'$. For this consider the perturbative expansion in the fields as in (\ref{taylorex}). One can show that if one assumes that the zeroth order KSEs
are imposed,  ({\ref{lichcorr}}) does not have an $\alpha'$ correction apart from the gaugino term, which leads to the condition
\begin{eqnarray}
\label{cond2}
{\slashed{\tilde{F}}} \chi_+ = 0+{\cal O}(\alpha') \ .
\end{eqnarray}
So  we cannot conclude that the KSEs, apart from the gaugino,
are implied from ({\ref{lichcorr}}) to order $\alpha'$. For this  some control over the $\alpha'^2$ terms
is required which  is not available. Observe that the above theorem also implies that all solutions of the gravitino and dilatino  KSEs are also solutions
 of the gaugino one. This is because  the modified Dirac-like operator $\mathscr{D}^{(+)}$ is constructed from only the gravitino and dilatino  KSEs but nevertheless
 the above theorem implies that the gaugino KSE is implied as well.


In the truncated theory, one can again formulate a Lichnerowicz type of theorem
provided that one imposes by hand the condition
\begin{eqnarray}
{\tilde{R}}_{\ell_1 \ell_2,mn} \Gamma^{\ell_1 \ell_2} \chi_+ =0	\ .
\end{eqnarray}
This condition (taking $0<q_1<{2 \over 7}$) is sufficient to ensure that the
RHS of ({\ref{lichcorr}}) can be written as a sum of positive definite terms, which must all vanish.

\newsection{A no-go theorem for AdS$_n$, $n \geq 4$ and $n=2$ backgrounds}

There are no AdS$_n$, $n \geq 4$ backgrounds in heterotic theory with or without $\alpha'$ corrections up to two loops in sigma model
perturbation theory.  This includes the case for which the theory is treated as exact up to and including  two loops.

The proof of this relies on the solution of the KSEs. Suppose that the fields are invariant under the symmetries of AdS$_n$.
Then we take a basis for the spacetime as $\{ {\mathbf{e}}^\lambda=
A {\bar{\mathbf{e}}}^\lambda, {\mathbf{e}}^i \}$
where ${\bar{\mathbf{e}}}^\lambda$ is a basis for $AdS_n$, and ${\mathbf{e}}^i$ is a basis for the internal space $M^{10-n}$.
We take $H$ to be a 3-form on $M^{10-n}$.
The components of $H$, and the conformal factor $A$, depend only on the
co-ordinates of $M^{10-n}$.

To proceed, consider the gravitino KSE along the AdS$_n$ frame directions, see also appendix E.
This has no contribution from the 3-form $H$, and can be rewritten as
\begin{eqnarray}
{\bar{\nabla}}_\lambda \epsilon -{1 \over 2} \Gamma_\lambda \partial_i A \Gamma^i \epsilon =0 \ .	
\end{eqnarray}
where ${\bar{\nabla}}$ denotes the Levi-Civita connection on $AdS_n$.
The integrability condition of this equation implies that
\begin{eqnarray}
\label{intaux}
\bigg(\Gamma^\lambda {\bar{R}}_{\mu \lambda} + (1-n) (dA)^2 \Gamma_\mu
\bigg) \epsilon =0 \ .	
\end{eqnarray}
where ${\bar{R}}_{\mu \nu}$ is the Ricci tensor of ${\bar{\nabla}}$.
However, for $AdS_n$, ${\bar{R}}_{\mu \nu} = \kappa {\bar{g}}_{\mu \nu}$
where ${\bar{g}}$ is the metric on $AdS_n$, and $\kappa$ is a negative constant. The integrability condition ({\ref{intaux}}) is then
equivalent to
\begin{eqnarray}
\bigg( \kappa+ (1-n) (dA)^2 \bigg) \epsilon=0	
\end{eqnarray}
which admits no solution as $\kappa<0$ and $n \geq 4$.

The above argument clearly applies for all backgrounds with $dH=0$, and so excludes the existence of AdS$_n$, $n>3$, backgrounds for
the common sector and the heterotic theory for which there is not an anomalous correction to the Bianchi identity. This result is also valid
for the AdS$_n$, $n>3$ solutions of the truncated  theory as well. It remains to investigate the existence of AdS$_n$, $n>3$, backgrounds in perturbative heterotic theory with
an anomalous contribution to the Bianchi identity, $dH\not=0$. In this case, the argument above implies that at zeroth order in $\alpha'$, there are no such solutions. Furthermore, it also excludes
the existence of AdS$_n$, $n>3$,  solutions up and including two loops  in sigma model perturbation theory that preserve all the symmetries of AdS$_n$. However such solutions
cannot  completely be excluded in higher orders as it is not known how the KSEs and field equations are corrected. There is the possibility that one can start from another
background which is allowed at zeroth order which then gets corrected in perturbation theory to an AdS$_n$, $n>3$  solution. Although this cannot be excluded, it may be a rather remote
possibility. We conclude therefore that up to order ${\cal O}(\alpha'^2)$ in perturbation theory there are no AdS$_n$, $n>3$, solutions to heterotic theory.

It remains to investigate the existence of AdS$_2$ solutions. It is a consequence of the investigation of near horizon geometries in \cite{hethor} that if $dH=0$, there
are no AdS$_2$ solutions. 
It should be emphasized though that the exclusion of 
AdS$_2$ backgrounds requires the use of the maximum principle. Otherwise heterotic AdS$_2$ backgrounds do exist \cite{stromingerads2}.  This result extends up to order $\alpha'^2$ in perturbation theory as it is unlikely that one can start from a different zeroth
order background and correct it at one-loop approximation to an  AdS$_2$ background-though we do not have  a proof for this. The existence of AdS$_2$ solutions for the truncated
theories will be examined elsewhere.

\vskip 0.5cm
\noindent{\bf Acknowledgements} \vskip 0.1cm
\noindent
JG is supported by the STFC grant, ST/1004874/1.
GP is partially supported by the  STFC rolling grant ST/J002798/1.
\vskip 0.5cm

\setcounter{section}{0}\setcounter{equation}{0}

\appendix{Notation and conventions}

Our form conventions are as follows. Let $\omega$ be a k-form, then
\begin{eqnarray}
\omega={1\over k!} \omega_{i_1\dots i_k} dx^{i_1}\wedge\dots \wedge dx^{i_k}~,
\end{eqnarray}
and
\begin{eqnarray}
d\omega={1\over k!} \partial_{i_1} \omega_{i_2\dots i_{k+1}} dx^{i_1}\wedge\dots \wedge dx^{i_{k+1}}~,
\end{eqnarray}
leading to
\begin{eqnarray}
(d\omega)_{i_1\dots i_{k+1}}= (k+1) \partial_{[i_1} \omega_{i_2\dots i_{k+1}]}~.
\end{eqnarray}
Furthermore, we write
\begin{eqnarray}
\omega^2= \omega_{i_1\dots i_k} \omega^{i_1\dots i_k}~,~~~\omega^2_{i_1 i_{2}}=\omega_{i_1j_1\dots j_{k-1}} \omega_{i_2}{}^{j_1\dots j_{k-1}} \ .
\end{eqnarray}

It is well-known that for every form $\omega$, one can define a Clifford algebra element ${\slashed \omega}$ given by
\begin{eqnarray}
{\slashed\omega}=\omega_{i_1\dots i_k} \Gamma^{i_1\dots i_k}
\end{eqnarray}
where $\Gamma^i$, $i=1,\dots n$, are the Dirac gamma matrices. In addition we introduce the notation
\begin{eqnarray}
{\slashed\omega}_{i_1}= \omega_{i_1 i_2 \dots i_k} \Gamma^{i_2\dots i_k}~,~~~{\slashed \Gamma\mkern-4.0mu \omega}_{i_1}= \Gamma_{i_1}{}^{
i_2\dots i_{k+1}} \omega_{i_2\dots i_{k+1}}~.
\end{eqnarray}
The rest of our spinor conventions can be found in \cite{hetclas}

\appendix{Bianchi identities and field equations of the perturbed AdS$_3$ background}
\label{apb}

Using the conditions derived from the KSEs, the anomalous Bianchi identity and the field equations
up to order $\alpha'^2$ can be written as follows. The Bianchi identity is
\begin{eqnarray}
 dG &= -\frac{\alpha'}{4} \left[ \text{tr} \left( \check{R}^{(7)} \wedge \check{R}^{(7)} \right) - \text{tr} (F \wedge F) \right] +{\cal{O}}(\alpha'^2)~.
\end{eqnarray}
The field equations can also be expressed as,
\begin{eqnarray}
 \nabla^2 \Phi &=&  2 (d\Phi)^2 - \frac{1}{12} G^2 + \frac{1}{2} X^2
 \cr
 && \qquad + \frac{\alpha'}{16} \left[ \check{R}^{(7)}_{i j, k \ell} \check{R}^{(7) i j, k \ell} - F_{i j a b} F^{i j a b} \right]+{\cal{O}}(\alpha'^2)~,
 \cr
 \nabla^k G_{i j k} &=&  2 \partial^k \Phi G_{i j k}+{\cal{O}}(\alpha'^2)~,
\end{eqnarray}
and
\begin{eqnarray}
 R^{(7)}_{i j} = \frac{1}{4} G_{i k_1 k_2} G_j{}^{k_1 k_2} - 2 \nabla_i \nabla_j \Phi
  - \frac{\alpha'}{4} \left[ \check{R}^{(7)}_{i k, s t} \check{R}^{(7)}{}_j{}^{k, s t} - F_{i k a b} F_j{}^{k a b} \right] +{\cal{O}}(\alpha'^2) \ .
\end{eqnarray}
From the latter,   we find that
\begin{eqnarray}
 R^{(7)}
 &=& -\frac{6}{\ell^2} A^{-2}  + \frac{5}{12} G^2 + \frac{1}{2} X^2  - 4 (d\Phi)^2
 \cr
 && \qquad - \frac{3 \alpha'}{8} \left[ \check{R}^{(7)}_{i j, k \ell} \check{R}^{(7) i j, k \ell} - F_{i j a b} F^{i j a b} \right]+{\cal{O}}(\alpha'^2) \ .
\end{eqnarray}
This is used in  the proof of the Lichnerowicz-type theorem for backgrounds with $\alpha'$ corrections.

\appendix{Dilaton field equation}

There are two ways to derive the dilaton field equation. The first
makes use only of the bosonic field equations and yields the covariant derivative of the dilaton field equation. As a result
the dilaton field equation is determined from the other field equations up to a constant. The second method
makes use of the integrability conditions of the dilatino and gravitino
KSE, together with the gaugino KSE and the bosonic field equations, and yields the
dilaton field equation. In this Appendix we present both methods.

\subsection{Dilaton equation via bosonic conditions alone.}

In the absence of an anomaly contribution, the dilaton field equation is implied by the field equations of the other fields, up to an additive constant \cite{tseyt}.
We shall re-examine this in the presence of an anomalous contribution both in perturbation theory and in the context of the truncated theory.

Taking the divergence of the Einstein equation (\ref{ealpha}), one has
\begin{equation}
 \nabla^M R_{M N} = \frac{1}{4} \nabla^M H^2_{M N} - 2 \nabla^2 \nabla_N \Phi - \frac{\alpha'}{4} \nabla^M \left[ \tilde{R}_{M K, S T} \tilde{R}_N{}^{K, S T} - F_{M K a b} F_N{}^{K a b} \right]~.
 \label{dive}
\end{equation}
Then observe that
\begin{align}
 -2 \nabla^2 \nabla_N \Phi &= -2 \nabla_N \nabla^2 \Phi - 2 R_{M N} \nabla^M \Phi
 \nonumber \\
 &= -2 \nabla_N \nabla^2 \Phi + 2 \nabla_N (d\Phi)^2 - \frac{1}{2} H^2_{M N} \nabla^M \Phi
 \nonumber \\
 & \qquad + \frac{\alpha'}{2} \left[ \tilde{R}_{M K, S T} \tilde{R}_N{}^{K, S T} - F_{M K a b} F_N{}^{K a b} \right] \nabla^M \Phi~,
\end{align}
where the Einstein equation has been used again.
Substituting this back into (\ref{dive}),
\begin{align}
 \nabla^M R_{M N} &= -2 \nabla_N \nabla^2 \Phi + 2 \nabla_N (d\Phi)^2 + \frac{1}{4} e^{2\Phi} \nabla^M ( e^{-2\Phi} H^2_{M N} )
 \nonumber \\
 & \qquad - \frac{\alpha'}{4} \nabla^M ( e^{-2\Phi} \left[ \tilde{R}_{M K, S T} \tilde{R}_N{}^{K, S T} - F_{M R a b} F_N{}^{R a b} \right] )~.
\end{align}
On the other hand, the divergence of the Ricci curvature is
\begin{align}
 \nabla^M R_{M N} &= \frac{1}{2} \nabla_N R
 \nonumber \\
 &= \nabla_N \left( \frac{1}{8} H^2 - \nabla^2 \Phi - \frac{\alpha'}{8} \left[ \tilde{R}_{M N, S T} \tilde{R}^{M N, S T} - F_{M N a b} F^{M N a b} \right] \right)
\end{align}
so, rearranging the Einstein equation
\begin{align}
 \nabla_N \nabla^2 \Phi &= 2 \nabla_N (d\Phi)^2 + \frac{1}{4} e^{2\Phi} \nabla^M ( e^{-2\Phi} H^2_{M N} ) - \frac{1}{8} \nabla_N H^2
 \nonumber \\
 & \qquad - \frac{\alpha'}{4} e^{2\Phi} \nabla^M ( e^{-2\Phi} \left[ \tilde{R}_{M K, S T} \tilde{R}_N{}^{K, S T} - F_{M K a b} F_N{}^{K a b} \right] )
 \nonumber \\
 & \qquad + \frac{\alpha'}{8} \nabla_N \left[ \tilde{R}_{L_1 L_2, L_3 L_4} \tilde{R}^{L_1 L_2, L_3 L_4} - F_{L_1 L_2 a b} F^{L_1 L_2 a b} \right]~.
 \label{dddphi}
\end{align}
From the Bianchi identity for $H$,
\begin{align}
 H_{M K L} dH^{M N K L} &\equiv 3 H_{M K L} \nabla^M H^{N K L} - \frac{1}{2} \nabla^N H^2
 \nonumber \\
 &= \frac{3\alpha'}{2} H_{M K L} \left[ \tilde{R}^{M N,}{}_{S T} \tilde{R}^{K L, S T} - F^{M N}{}_{a b} F^{K L a b} \right]~.
 \end{align}
 Thus
 \begin{align}
 H_{M K L} \nabla^M H^{N K L} &= \frac{1}{6} \nabla^N H^2 + \frac{\alpha'}{2} H_{M K L} \left[ \tilde{R}^{M N,}{}_{S T} \tilde{R}^{K L, S T} - F^{M N}{}_{a b} F^{K L a b} \right]~.
\end{align}
This together with the field equation for $H$ gives
\begin{align}
 \frac{1}{4} e^{2\Phi} \nabla^M ( e^{-2\Phi} H^2_{M N} ) &= \frac{1}{4} e^{2\Phi} \nabla^M ( e^{-2\Phi} H_{M K L} ) H_N{}^{K L} + \frac{1}{4} H_{M K L} \nabla^M H_N{}^{K L}
 \nonumber \\
 &= \frac{1}{24} \nabla^N H^2 + \frac{\alpha'}{8} H_{M K L} \left[ \tilde{R}^M{}_{N, S T} \tilde{R}^{K L, S T} - F^M{}_{N a b} F^{K L a b} \right]~.
\end{align}
Substituting this back into (\ref{dddphi}), we find
\begin{align}
 \nabla_N \nabla^2 \Phi &= 2 \nabla_N (d\Phi)^2 - \frac{1}{12} \nabla_N H^2 + \frac{\alpha'}{8} H_{M K L} \left[ \tilde{R}^M{}_{N, S T} \tilde{R}^{K L, S T} - F^M{}_{N a b} F^{K L a b} \right]
 \nonumber \\
 & \qquad - \frac{\alpha'}{4} e^{2\Phi} \nabla^M ( e^{-2\Phi} \left[ \tilde{R}_{M K S T} \tilde{R}_N{}^{K S T} - F_{M K a b} F_N{}^{K a b} \right] )
 \nonumber \\
 & \qquad + \frac{\alpha'}{8} \nabla_N \left[ \tilde{R}_{L_1 L_2 L_3 L_4} \tilde{R}^{L_1 L_2 L_3 L_4} - F_{L_1 L_2 a b} F^{L_1 L_2 a b} \right]~.
 \label{dddphix}
\end{align}
Clearly this implies the field equation of the dilaton up to a constant for theories that do not have an anomaly contribution.

For the theories with an anomaly contribution, the derivation requires further examination. First, let us focus on the quadratic terms in $F$ in (\ref{dddphix}).
From the Bianchi identity for $F$,
\begin{equation}
 0 = \frac{3}{2} F_{M K} \big(\nabla^{[M} F^{NK]} + [A^{[M}, F^{NK]}] \big)  = F_{M K} \big(\nabla^M F^{N K}+[A^M,F^{NK}] \big) - \frac{1}{4} \nabla^N F^2~,
\end{equation}
where the gauge indices, which are traced over in all terms quadratic in $F$, have been suppressed.
Furthermore, using the field equation for $F$, one has
\begin{align}
 \frac{\alpha'}{4} e^{2\Phi} \nabla^M ( e^{-2\Phi} F_{M K} F_N{}^K ) &= \frac{\alpha'}{4} e^{2\Phi} \nabla^M ( e^{-2\Phi} F_{M K} ) F_N{}^K + \frac{\alpha'}{4} F_{M K} \nabla^M F_N{}^K
 \nonumber \\
 &= \frac{\alpha'}{8} H^M{}_K{}^L F_{M L} F_N{}^K + \frac{\alpha'}{16} \nabla_N F^2~.
\end{align}
Therefore the terms quadratic in $F$ sum in (\ref{dddphix}) to  just $-\frac{\alpha'}{16} \nabla_N F^2$ as expected.

It remains to see what happens to the quadratic terms in $\tilde R$ in (\ref{dddphix}).
First, consider the perturbative case, with ${\tilde{R}}={\check{R}}$.
The curvature terms can be most straightforwardly simplified by making use
of the appropriate curvature Bianchi identity, given by
\begin{align}
{\check{\nabla}}_{[K} {\check{R}}_{MN],SL}=-H^Q{}_{[MN} {\check{R}}_{K]Q,SL}
\end{align}
or equivalently,
\begin{align}
\label{curvcondaux1}
\nabla_{[K} 	 {\check{R}}_{MN],SL} = {1 \over 2} H^Q{}_{S[K} {\check{R}}_{MN],SL} -{1 \over 2} H^Q{}_{L[K} {\check{R}}_{MN],QS} \ .
\end{align}
On contracting the above identity over the $K,S$ indices, we
obtain the following useful expression, valid to zeroth order in $\alpha'$:
\begin{align}
\label{curvcondaux2}
\nabla^K \bigg(e^{-2 \Phi} {\check{R}}_{MN,KL}\bigg)
-e^{-2 \Phi} H^{QK}{}_{[M} {\check{R}}_{N]K,QL}
-{1 \over 2} e^{-2 \Phi} H^{QK}{}_L {\check{R}}_{MN,QK}=0	
\end{align}
where we have made use of the identity
\begin{align}
{\check{R}}_{NL} \equiv {\check{R}}^K{}_{N,KL}=-2 \nabla_N \nabla_L \Phi
- \partial^Q \Phi H_{QNL} +{\cal{O}}(\alpha')	
\end{align}
which follows from the Einstein equations ({\ref{ealpha}}), and the
gauge field equations for $H$.
On using ({\ref{curvcondaux1}}) and ({\ref{curvcondaux2}}), it follows that
\begin{align}
{1 \over 8} H_{Q_1 Q_2 Q_3} {\check{R}}^{Q_1}{}_{N,L_1 L_2}
{\check{R}}^{Q_2 Q_3, L_1 L_2}
-{1 \over 4} e^{2 \Phi} \nabla^M \bigg(e^{-2 \Phi} {\check{R}}_{ML_1, L_2 L_3} {\check{R}}_{N}{}^{L_1, L_2 L_3} \bigg)	
\nonumber \\
=-{1 \over 16} \nabla_N \bigg({\check{R}}_{L_1 L_2, L_3 L_4}
{\check{R}}^{L_1 L_2, L_3 L_4} \bigg) + {\cal{O}}(\alpha') .
\end{align}
The above expression produces the appropriate curvature squared term
in the dilaton equation at first order in $\alpha'$.
For the truncated theory,  the appropriate curvature term
in the dilaton equation is obtained  if
({\ref{curvcondaux1}}) and ({\ref{curvcondaux2}}) hold exactly, on replacing
${\check{R}}$ with ${\tilde{R}}$.

\subsection{Dilaton equation via supersymmetry}

Another way to obtain the dilaton field equation is to assume the KSEs, see also \cite{hetclas}.
In particular, consider the condition
\begin{align}
{\slashed{\nabla}} \bigg( \big(\slashed{\partial}\Phi-{1\over12} \slashed{H}\big)\epsilon \bigg) =0+{\cal O}(\alpha'^2) \ .
\end{align}

On expanding out the LHS, one obtains, after making use of the $H$ gauge field equation, and the gravitino and dilatino KSE the following expression:
\begin{align}
\bigg(\nabla^2 \Phi -2 (d\Phi)^2 +{1 \over 12}H^2
-{1 \over 48} {\slashed{dH}} \bigg) \epsilon =0+{\cal O}(\alpha'^2) \ .
\end{align}
The $dH$ bianchi identity term can then be rewritten to imply
\begin{align}
\label{dilatoneqsusy}
\bigg(\nabla^2 \Phi -2 (d\Phi)^2 +{1 \over 12}H^2
-{\alpha' \over 16} \big({\tilde{R}}_{L_1 L_2, L_3 L_4} {\tilde{R}}^{L_1 L_2, L_3 L_4}-F_{L_1 L_2 ab} F^{L_1 L_2 ab} \big)
\nonumber \\
+{\alpha' \over 32} {\tilde{R}}_{N_1 N_2, ST} \Gamma^{N_1 N_2}
{\tilde{R}}_{M_1 M_2,}{}^{ST} \Gamma^{M_1 M_2}
- {\alpha' \over 32} {\slashed{F}}_{ab} {\slashed{F}}^{ab}
 \bigg) \epsilon=0+{\cal O}(\alpha'^2) \ .
\end{align}

In the perturbative case, the final term on the LHS can be neglected as a consequence of the
gaugino KSE. The penultimate term can also be neglected because,
if ${\tilde{R}}={\check{R}}$ then
\begin{align}
{\tilde{R}}_{M_1 M_2, ST} \Gamma^{M_1 M_2} \epsilon
= {\hat{R}}_{ST, M_1 M_2} \Gamma^{M_1 M_2} \epsilon + {\cal{O}}(\alpha')
= {\cal{O}} (\alpha')
\end{align}
as a consequence of the integrability conditions of the gravitino KSE.
The first line  of ({\ref{dilatoneqsusy}}) then implies the dilaton
equation to the required order, provided that the zeroth order in $\alpha'$
part of the Killing spinor $\epsilon$ is nonzero.

In the truncated case, again the final term on the LHS of ({\ref{dilatoneqsusy}}) vanishes due to the gaugino KSE. The penultimate
term vanishes if one assumes that
\begin{align}
{\tilde{R}}_{M_1 M_2, ST} \Gamma^{M_1 M_2} \epsilon =0
\end{align}
and then ({\ref{dilatoneqsusy}}) implies the dilaton equation.

\appendix{Geometry of AdS$_3$ backgrounds and $\tau$ spinors}

 To investigate the conditions on the geometry of AdS$_3$ backgrounds
imposed by the  $\tau_\pm$ Killing spinors, we note that there are two
linearly independent spinors given by
\begin{eqnarray}
\epsilon^1=e^{-{z\over\ell}} \tau_+-\ell^{-1} r A^{-1} e^{-{z\over\ell}}\Gamma_{-z} \tau_+~,~~~\epsilon^2= e^{{z\over\ell}} \tau_-~.
\end{eqnarray}
Furthermore, as $\tau_-=A \Gamma_{-z} \tau_+$, the two Killing spinors can be rewritten as
\begin{eqnarray}
\epsilon^1=e^{-{z\over\ell}} \tau_+-\ell^{-1} r A^{-1} e^{-{z\over\ell}}\Gamma_{-z} \tau_+~,~~~\epsilon^2= e^{{z\over\ell}} A \Gamma_{-z} \tau_+~.
\end{eqnarray}
Next one can compute the 1-form Killing spinor bilinears  which after an appropriate normalization can be written as
\begin{eqnarray}
 \kappa^+ &=& e^{\frac{2 z}{\ell}} {\bf{e}}^+~,~~~
 \kappa^- = e^{-\frac{2 z}{\ell}} \left( {\bf{e}}^- - 2 \ell^{-2} r^2 A^{-2} {\bf{e}}^+ + 2 \ell^{-1} r dz \right)~,
 \cr
 \kappa^z &=& dz - 2 \ell^{-1} r A^{-2} {\bf{e}}^+~.
\end{eqnarray}
The Lie bracket algebra of the associated vector fields is
\begin{equation}
 [ \kappa^+, \kappa^z ] = -2 \ell^{-1} A^{-2} \kappa^+ \qquad [ \kappa^-, \kappa^z ] = 2 \ell^{-1} A^{-2} \kappa^- \qquad [ \kappa^+, \kappa^- ] = 2 \ell^{-1} \kappa^z ,
\end{equation}
which is isomorphic to $\mathfrak{sl}(2,{\mathbb{R}})$, where we have denoted the 1-forms and the associated vector fields with the same symbols.
Furthermore after some algebra one can show that the above vector fields commute with the vector fields constructed from the Killing spinors determined in terms of the $\sigma$'s.
This confirms the statement in section 2.2 that  the two choices of Killing spinors for AdS$_3$ backgrounds in terms of $\sigma$'s and $\tau$'s correspond to the two different choices of
parallelization of AdS$_3$ as a group manifold in terms of the left and right group actions.

\appendix{The no-go theorem for AdS$_n$, $n \not= 3$  backgrounds revisited}

The non-existence of  AdS$_n$, $n \geq 4$ backgrounds in heterotic theory with or without $\alpha'$ corrections up to two loops in sigma model
perturbation theory can be stated in a coordinate system similar to that used in the analysis of AdS$_3$ backgrounds as follows.
Suppose that the fields are invariant under the symmetries of AdS$_n$. Then  we have
\begin{eqnarray}
 ds^2 &=& 2 \mathbf{e}^+ \mathbf{e}^- + A^2 dz^2 + A^2 e^{-{2z\over\ell}} \sum^{n-3}_{a=1} (dx^a)^2+ds^2 \left( M^{10-n} \right) ,
 \cr
  H &=&  G ,
 \end{eqnarray}
The gravitino KSE along the AdS$_n$, $n>3$, directions reads
\begin{eqnarray}
 \partial_u \epsilon_\pm + A^{-1} \Gamma_{+ z} \left( \ell^{-1} - \Xi_- \right) \epsilon_\mp&=&0~,
 \cr
  \partial_r \epsilon_\pm - A^{-1} \Gamma_{- z} \Xi_+ \epsilon_\mp&=&0~,
  \cr
  \partial_z \epsilon_\pm - \Xi_\pm \epsilon_\pm + 2 r A^{-1} \Gamma_{- z} \Xi_+ \epsilon_\mp&=&0~,
 \cr
 \partial_a \epsilon_+ + A^{-1} \Gamma_{z a} \Xi_+ \epsilon_+&=&0~,
 \cr \partial_a \epsilon_- + A^{-1} \Gamma_{z a} \left( \Xi_- - \ell^{-1} \right) \epsilon_-&=&0~,
\end{eqnarray}
where $\Gamma_\pm\epsilon_\pm=0$ and
\begin{equation}
 \Xi_\pm = \mp \frac{1}{2 \ell} + \frac{1}{2} \slashed{\partial} A \Gamma_z~.
\end{equation}

Using the relations
\begin{eqnarray}
 \Xi_\pm \Gamma_{z +} + \Gamma_{z +} \Xi_\mp &=& 0~,~~~
 \Xi_\pm \Gamma_{z -} + \Gamma_{z -} \Xi_\mp = 0~,
 \cr
 \Xi_\pm \Gamma_{z a} + \Gamma_{z a} \Xi_\pm &=& \mp \ell^{-1} \Gamma_{z a}~,
\end{eqnarray}
we find that there is only one AdS-AdS integrability condition which reads
\begin{equation}
 \left( {\Xi_\pm}^2 \pm \frac{1}{\ell} \Xi_\pm \right) \epsilon_\pm = 0~.
\end{equation}
On the other hand
\begin{eqnarray}
 {\Xi_\pm}^2 &=& \left[ \mp \frac{1}{2 \ell} + \frac{1}{2} \slashed{\partial} A \Gamma_z \right] \left[ \mp \frac{1}{2 \ell} + \frac{1}{2} \slashed{\partial} A \Gamma_z \right]
 \cr
 &=& \frac{1}{4 \ell^2} \mp \frac{1}{2 \ell} \slashed{\partial} A \Gamma_z - \frac{1}{4} (dA)^2~,
 \end{eqnarray}
and so
 \begin{eqnarray}
 {\Xi_\pm}^2 \pm \frac{1}{\ell} \Xi_\pm &= -\frac{1}{4 \ell^2} - \frac{1}{4} (dA)^2~.
\end{eqnarray}
Therefore the integrability condition cannot be satisfied for $\ell < \infty$.  The rest of the analysis is as in section 4.

\end{document}